\documentclass[%
 preprint,
 amsmath,amssymb,
 aps, physrev,
]{revtex4-2}

\usepackage{graphicx}
\usepackage{dcolumn,comment,subfigure}
\usepackage{bm}
\usepackage{xcolor,comment}
\usepackage{hyperref,graphics}
\usepackage[mathlines]{lineno}
\usepackage{soul}

\begin{document}


\title{\textbf{A General Framework for Linking Free and Forced Fluctuations via Koopmanism} 
}%

\author{Valerio Lucarini\footnote{Email: \texttt{v.lucarini@leicester.ac.uk}}}
 \author{Manuel Santos Guti\'errez}%
\author{John Moroney}%
\affiliation{%
School of Computing and Mathematical Sciences, University of Leicester, Leicester, UK
}%

\author{Niccol\`o Zagli}
\affiliation{NORDITA, Stockholm, Sweden
}%

\date{\today}

\begin{abstract}
The link between forced and free fluctuations for nonequilibrium systems can be described via a generalized version of the celebrated fluctuation-dissipation theorem. The use of the formalism of the Koopman operator makes it possible to deliver an intepretable form of the response operators written as a sum of exponentially decaying terms, each associated one-to-one with a mode of natural variability of the system. Here we showcase on a stochastically forced version of the celebrated Lorenz '63 model the feasibility and skill of such an approach by considering different Koopman dictionaries, which allows us to treat also seamlessly coarse-graining approaches like the Ulam method. Our findings provide support for the development of response theory-based investigation methods also in an equation-agnostic, data-driven environment. 
\end{abstract}

\maketitle

\section{Introduction}
One of the key challenges in science is to be able to predict some desired yet still unobserved properties of a system of interest on the basis of what is currently known about such a system \cite{Sarewitz1999}. In the case of a complex system, rather than focusing on an individual trajectory, we are often interested in evaluating the statistical impact of adding some extra forcing. In this scenario, prediction amounts to being able to anticipate the features of the forced fluctuations of the system on the basis of what we know on the unforced, or naturally-occurring, ones. For systems near thermodynamic equilibrium, the fluctuation-dissipation theorem (FDT) quantifies the link between forced and free fluctuations for a weakly forced system \cite{Kubo.1966}. Indeed, the response operators enabling the computation of the (in general) time-dependent response of the system to forcings are cast as rather intuitive correlation functions evaluated for the unperturbed system. The FDT has been extended to nonequilibrium systems \cite{Marconi2008,Baiesi2013,Sarracino2019,Lucarini2020PRSA} as well as towards representing nonlinear effects \cite{LucariniColangeli2012,Colangeli_2014}. A key assumption here is that the invariant measure of the unperturbed system is absolutely continuous with respect to the Lebesgue measure of the phase space of the system, as is realized, e.g., in the case of (hypo)elliptic diffusion processes \cite{pavliotisbook2014}. In the case of dissipative chaotic systems, whilst the FDT in its usual form cannot be readily applied, it is possible to rigorously develop a  response theory \cite{Ruelle1997DifferentiationSRBstates,ruelle_nonequilibrium_1998,Dolgopyat2004} that can be harnessed to predict the change in the statistics due to perturbations \cite{reicklinear2002,Cessac2007,Lucarini2009DispersionNonLinearResponseLorenz,LucariniSarno2011}.

Although the formal properties of the response formulas allow for a straightforward treatment of experimental data, even when coming from high-dimensional, multiscale systems - see discussion in \cite{Lucarini2017} - it is extremely challenging, in practice, to compute the response operators directly from the evolution equations of the system. Whilst response formulas are (somewhat) deceptively simple, implementing stable and accurate algorithms is far from easy \cite{MajdaAbramov2008,MajdaAbramov2010,Cooper2011,Cooper2013,Wang2013,Chandramoorthy2020,Angxiu2021,Angxiu2023}.  

A possible workaround to address these difficulties comes from the use of a combination of theory-informed and data-driven methods. Giorgini and collaborators have recently proposed to estimate response functions using score-based generative modelling \cite{Giorgini2024,Giorgini2025}, which eases the need for achieving an accurate estimation of the invariant measure of the unperturbed system, which is instead needed in the case of Gaussian mixture statistical models  \cite{Cooper2011,Cooper2013}. Whilst the generative modelling is indeed extremely promising and has been shown to work accurately in some illustrative cases, it has also been emphasized that challenges emerge in the data training process \cite{Giorgini2024,Giorgini2025}.   

Finite-state and finite-time Markov chains \cite{Norris1998} are stochastic processes that are widely used to represent at coarse-grained level the evolution of possibly highly complex systems, whereby a finite region of a phase space of the system is associated with a state of the chain and a stochastic matrix describes the probability of transition between the states for the chosen time-step \cite{Gaspard2004,attal2010markov}. This viewpoint has considerable relevance in terms of data-driven analysis of complex systems according to the Markov state modeling (MSM) strategy \cite{Laio2006,Pande2010,Bowman2014,Husic2018}. Following earlier  results \cite{Seneta1993,Mitrophanov_2024}, it has recently been possible to derive closed formulas, easily amenable to implementation in current computational software, describing the response of Markov chains to perturbations \cite{Lucarini2016MarkovChains,Antown2018,SantosJSP}. This has led to the proposal for an equation-agnostic, fully data-driven response theory based on Markov chains \cite{Lucarini2025}.

In the last couple of decades, Koopmanism has emerged as a very powerful angle for studying at the evolution of possibly complex nonlinear systems - both stochastic and deterministic ones - by studying the infinite-dimensional linear operators  describing the evolution of observables \cite{Mezic2005,Budinisic2012}. Koopmanism has the great advantage of supporting a purely data-driven analysis of systems via variants of the extended dynamical mode decomposition (eDMD) algorithms  \cite{brunton2022data,Colbrook2024Multi}. In \cite{Santos2022,Chekroun2024} we showed how to express the response operators of a general class of stochastic dynamical systems as a sum of terms, each associated with a natural mode of variability described by an eigenmode identified by the Koopman analysis. This angle has the advantage of adding a layer of interpretability to the FDT and of providing a clear condition - the closure of the spectral gap - underpinning the occurrence of critical transitions or tipping points \cite{LucariniChekroun2023,LucariniChekroun2024}. Additionally, and very importantly, it take{\color{red}s} full advantage of the eDMD algorithms, as shown for some one and two-dimensional systems in  \cite{Zagli2024}. 

In this contribution we want to provide a unifying perspective linking the recent results presented in \cite{Zagli2024} and \cite{Lucarini2025}. The goal here is to investigate the response and fluctuations of the classical Lorenz '63 model \cite{lorenzdeterministic1963} supplemented with weak non-degenerate and isotropic additive noise, which guarantees the existence of a unique absolutely continuous invariant measure with respect to Lebesgue \cite{Bolley2012,pavliotisbook2014}. We will compute response operators for such a system using the Koopman formalism where we first select as functional dictionary  characteristic functions of specific subsets of the phase space. This approach is equivalent to constructing a Markov chain coarse-grained representation of the  continuous dynamics \cite{Lucarini2025}. Along these lines, we will construct the so-called Ulam partition  \cite{Ulam1960} considering 1) a regular gridding of the phase space in cubes of identical size, and 2) an irregular, data-driven  gridding in Vorono\"i cells \cite{Aurenhammer1991}  constructed by performing k-means clustering
\cite{Forgy65,Lloyd82} of the trajectories of the system, in the spirit of MSM. Additionally, we will consider as dictionary functions a suitable set of {tensorized Chebyshev polynomials} \cite{Abramowitz1964} in the three variables of the system.

The goal of this study is two-fold. First, we compare the response functions obtained using the different aforementioned methodologies and show the duality between Koopman/Markov semigroup formulas and the FDT. Secondly, we assess the convergence of the response operators as we consider finer partitions of the phase space  or {higher degrees of polynomials}. This goes well beyond testing the convergence of the invariant measure of the reduced order models, which is the subject of the Ulam conjecture \cite{Ulam1960,Froyland1998}. We will also discuss to what extent it is possible to approximate the response operators by considering only few dominant modes of variability, hence targeting the possibility of a reduced order modelling of the  response of the system. The paper is structured as follows. In Sect. \ref{model} we introduce the specific model we investigate in our study. In Sect. \ref{koopman} we provide a brief overview of the key aspects of the Koopman operator-based analysis of the output of the model, in its reference state and in the case perturbations are applied. In Sect. \ref{responseoperators} we show the performance of the methods proposed here to predict the response of the system to prototypical perturbations.  In Sect. \ref{conclusions} we present our conclusions and perspectives for future work.


\section{The Model}\label{model}
Our analysis is performed on the following elliptic diffusion process $\mathrm{d}\mathbf{x}=\mathbf{F}(\mathbf{x})\mathrm{d}t+\Sigma\mathrm{d}\mathbf{W}$ of the form:
    \begin{equation}\label{Themodel}
     \begin{split}   
     \mathrm{d}x&=\sigma(y-x)\mathrm{d}t+\epsilon\mathrm{d}W_1\\
\mathrm{d}y&=x(r-z)\mathrm{d}t-y\mathrm{d}t+\epsilon\mathrm{d}W_2\\     \mathrm{d}z&=xy\mathrm{d}t-\beta z\mathrm{d}t+\epsilon\mathrm{d}W_3
    \end{split}
\end{equation}
where the drift is given by the vector field of the Lorenz '63 system \cite{lorenzdeterministic1963},  $\mathrm{d}W_j$, $j=1,2,3$ are increments of independent Wiener processes, and $\epsilon>0$. We choose as reference parameters $\sigma=10$, $r=28$, $\beta=8/3$, which lead to chaotic dynamics for the deterministic version of the model \cite{Tucker2002}, and select $\epsilon=2$. The model is integrated with the standard Euler-Maruyama scheme \cite{kloeden_1992} with a small time-step $dt=0.001$, well within the range of convergence.  

The density of the probability distribution of the system $\rho(\mathbf{x},t)$ evolves according to the Fokker-Planck equation 
\begin{equation}\label{FPE}
\partial_t\rho(\mathbf{x},t)=\mathcal{L}_0\rho(\mathbf{x},t)=-\nabla\cdot(\mathbf{F}(\mathbf{x})\rho(\mathbf{x},t))+\frac{\epsilon^2}{2}\Delta\rho(\mathbf{x},t)
\end{equation}
where $\mathbf{F}$ is the drift term, $\nabla \cdot$ is the divergence operator, $\Delta$ is the Laplacian operator. 

The solutions to this equation constitute a strongly continuous one-parameter semigroup ($C_0$-semigroup) $\{\exp{(\mathcal{L}_0t)}\}_{t\geq0}$ where $\mathcal{L}_0$ is its infinitesimal generator densely defined in a suitable Banach space $\mathcal{B}$ \cite{pavliotisbook2014,chekroun2019c}. We have $\rho(\mathbf{x},t)=\Pi_0^t\rho(\mathbf{x},0)=\exp(\mathcal{L}_0t)\rho(\mathbf{x},0)$ where $\Pi_0^t$ is the Perron-Frobenius operator \cite{pavliotisbook2014} evolving forward the measure by a time interval $t$. The presence of non-degenerate noise and the contractivity of the drift  guarantees the existence of a unique invariant measure that is absolutely continuous with respect to Lebesgue, so that it can be written as $\rho_0(\mathbf{x})\mathrm{d}\mathbf{x}$, where $\mathcal{L}_0\rho_0(\mathbf{x})=0$ $\forall \mathbf{x}\in\mathbb{R}^3$ \cite{Bolley2012,pavliotisbook2014,ChekrounJSPI}.  Hence, it is natural to choose ${L}^2(\mathbb{R}^3,\rho_0(\mathbf{x})\mathrm{d}\mathbf{x})$ ($L^2_{\rho_0}$, in short) as the reference functional space.

Whilst $\rho_0(\mathbf{x},t)>0$ for every $\mathbf{x}$ in $\mathbb{R}^3$ as a result of the diffusion,  the measure is highly concentrated around the compact attractor of the deterministic system realized for $\epsilon=0$ because we are considering a weak noise; see related comment below. Indeed, Graham's field theory \cite{GrahamTel} indicates that the invariant measure can be described according to a large deviation law, so that  $\rho_0(\mathbf{x})\propto\exp(-2V(\mathbf{x})/\epsilon^2)$, where $V(\mathbf{x})$ is the so-called quasi-potential, which has a minimum on the support of the attractor. 

\section{Koopman Analysis}\label{koopman}

The generator of the backward Kolmogorov equation $\mathcal{K}_0$ acts on observables $f$ in $\mathcal{B^{\ast}}$ and is defined as the adjoint of $\mathcal{L}_0$: $\mathcal{K}_0f(\mathbf{x})=\mathcal{L}_0^\ast f(\mathbf{x})=\mathbf{F}(\mathbf{x})\cdot \nabla f(\mathbf{x})+\epsilon^2  \Delta f(\mathbf{x})/2$ where $\nabla$ indicates the gradient operator. The solutions of the induced partial differential equation yield a $C_0$-semigroup. We refer to $\mathcal{K}_0$ as the Koopman operator, even if this terminology is usually adopted only for the deterministic $\epsilon=0$ case) \cite{Budinisic2012}. Note that, in applications, the Koopman operator or the Perron-Frobenius operator are constructed up to a specified sampling time $\mathrm{d}\tau$, resulting in a time discretization. Assuming no loss of semigroup property, $e^{\mathcal{L}_0^{\ast}t} = \left[ e^{\mathcal{L}_0^{\ast}\mathrm{d}\tau} \right]^{q}=\left[ \Pi^{\mathrm{d}\tau}_0 \right]^q$, provided that $t = q \times \mathrm{d}\tau$. 


As the drift is dissipative (i.e. it allows for a Lyapunov function) \cite{sparrow} and the noise is non-degenerate, we can safely exclude the singular spectrum from the spectral analysis of  $\mathcal{L}_0^{\ast}$ in 
$L^2_{\rho_0}$, and we can focus on the discrete spectrum only  \cite{engel2000,chekroun2019c}. 
We refer to the eigenvalues as $\{ \lambda_j \}_{j=0}^\infty$, where $\mathcal{R}(\lambda_i)\geq \mathcal{R}(\lambda_j)$ if $j\geq i$, and to the corresponding eigenfunctions $\{ \varphi_j \}_{j=0}^\infty$ such that $\mathcal{L}_0^{\ast}\varphi_j=\lambda_j\varphi_j$. We equivalently have the eigenfunctions of the Perron-Frobenius operator $\{\rho_j \}_{j=0}^\infty$ such that $\mathcal{L}_0\rho_j=\lambda_j\rho_j$, where  $\lambda_0=0$ and $\rho_0$ defines the unique invariant density. Correspondingly, $\varphi_0=const$. The system is exponentially mixing, so that $\mathcal{R}(\lambda_1)<0$. 

 Note that in what follows we omit the possible presence of geometric multiplicities of the eigenvalues, which would lead to the presence of polynomial terms in the spectral decomposition \cite{Santos2022}.

In practical terms a meaningful subset  of eigenvalues $\{ \lambda_j \}_{j=0}^{N-1}$ and eigenfunctions $\{ \varphi_j \}_{j=0}^{N-1}$ of the Koopman operator are estimated projecting the dynamics onto a finite basis of dictionary functions $\{\psi_k\}_{k=1}^N$. Specifically, 
we follow here \cite{Kutz2016,Zagli2024,Colbrook2024Multi} and define a set of dictionary functions $\{\psi_j \}_{j=1}^N$ spanning an $N-$dimensional subspace $\mathcal{F}_N \subseteq L^2_{\rho_0}$. 
We arrange the dictionary of variables as a vector-valued function $\boldsymbol{\Psi}: \mathcal{M} \to \mathbb{C}^{1\times N}$ 
\begin{equation}
    \boldsymbol{\Psi}(\mathbf{x}) = 
    \begin{pmatrix}
        \psi_1(\mathbf{x}) , \psi_2(\mathbf{x}) , \dots , \psi_N(\mathbf{x})
    \end{pmatrix}.
\end{equation}
Extended Dynamical Mode Decomposition (eDMD) \cite{Colbrook2024Multi} provides the best least square approximation of the representation matrix of the Koopman operator in the finite dimensional subspace $\mathcal{F}_N$ as $
    \mathbf{K} = \mathbf{G}^+\mathbf{A} \in \mathbb{C}^{N\times N}$ where $\mathbf{G}^+$ is the pseudo-inverse of the Gram matrix $\mathbf{G}$ and
\begin{equation}
    \mathbf{G} = \frac{1}{M}\sum_{i=1}^M \boldsymbol{\Psi}^*(\mathbf{x}_i) \boldsymbol{\Psi}(\mathbf{x_{i}})
\quad 
  \mathbf{A} = \frac{1}{M1}\sum_{i=1}^{M-1} \boldsymbol{\Psi}^*(\mathbf{x}_i) \boldsymbol{\Psi}(\mathbf{x_{i+1}}).\label{matrixA}
\end{equation}
where $M$ (plus 1) is the total number of observations $\mathbf{x}_i$ taken with the time interval $\mathrm{d}\tau$ and $\mathbf{\Psi}^*$ is the adjoint of $\mathbf{\Psi}$. The Spectral Mapping Theorem imposes that that the eigenvalues of $\mathcal{K}_0=\mathcal{L}^*_0$ can be written as 
    $\lambda_k= \ln \nu_k/\mathrm{d}\tau$,
where $\nu_j$ are the eigenvalues of the matrix $\mathbf{K}$ \cite{engel2000}.  Additionally, the eigenfunctions of $\mathcal{L}^*_0$ are estimated as 
\begin{equation}
\label{eq: eigenfunction}
    \varphi_k(\mathbf{x}) = \boldsymbol{\Psi}(\mathbf{x})\boldsymbol{\xi}_k,
\end{equation}
where $\boldsymbol{\xi}_k$ are the right eigenvectors of  $\mathbf{K}$. By performing a change of basis, we can express a general observable that can be written using the chosen  dictionary $f(\mathbf{x}) = \sum_k f_k \psi_k(\mathbf{x})$ also as a linear combination of the Koopman eigenfunctions: $ f(\mathbf{x}) = \sum_k \tilde{f}_k \varphi_k(\mathbf{x})$, where the $f_k$'s and the $\tilde{f}_k$'s are linked by a linear transformation \cite{Zagli2024}.

\begin{figure*}[t] 
    \centering
    \subfigure[ Statistics of the occupancy $\phi_i$ for corresponding regular cubic and Vorono\"i tessellations.]{
        \includegraphics[width=0.48\linewidth]{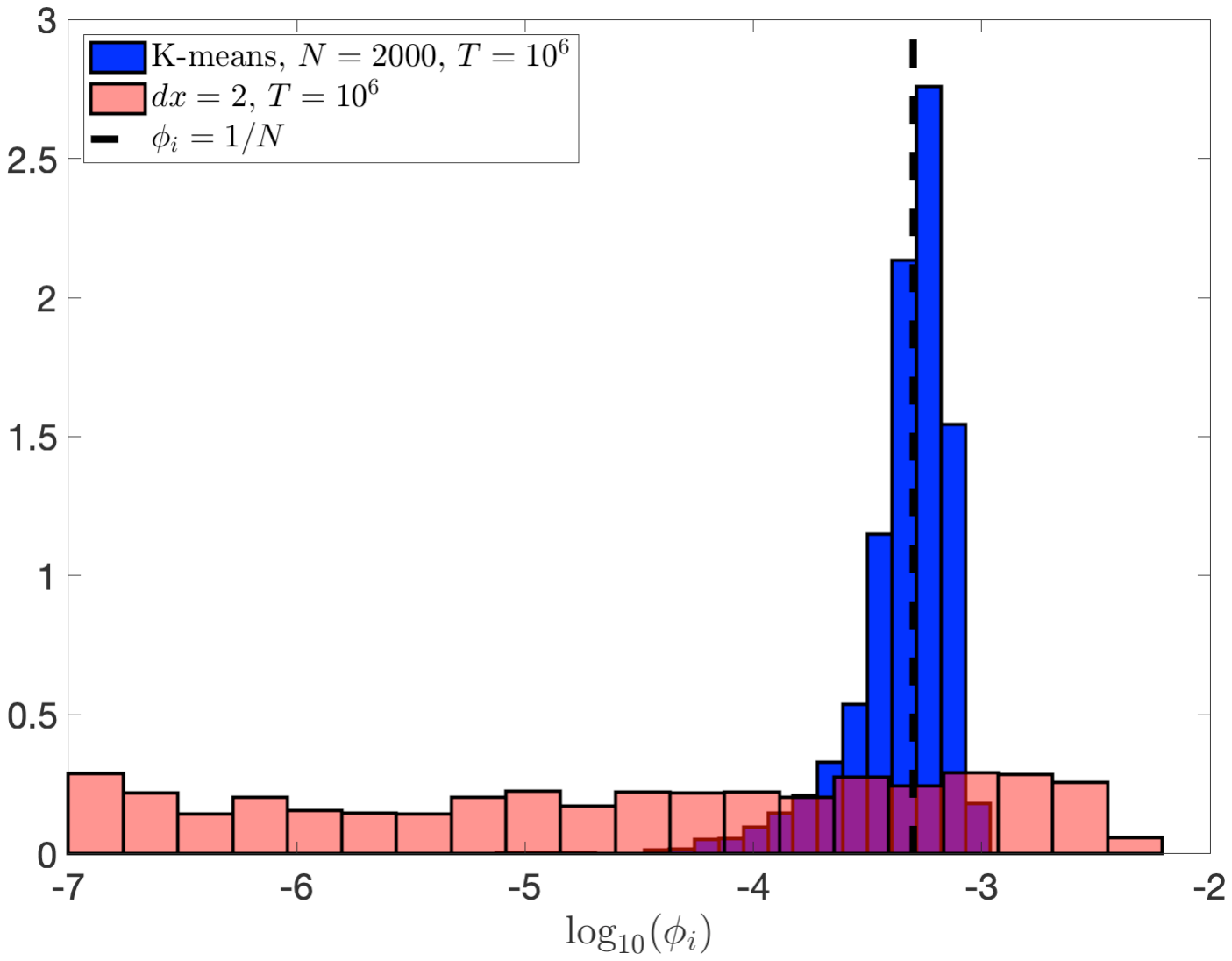}
        \label{fig:spectra_d} 
    }
    \hfill
    \subfigure[ Leading $\mathcal{L}_0$ eigenvalues via Ulam partition obtained with identical cubes of size $\mathrm{dx}$.]{
        \includegraphics[width=0.48\linewidth]{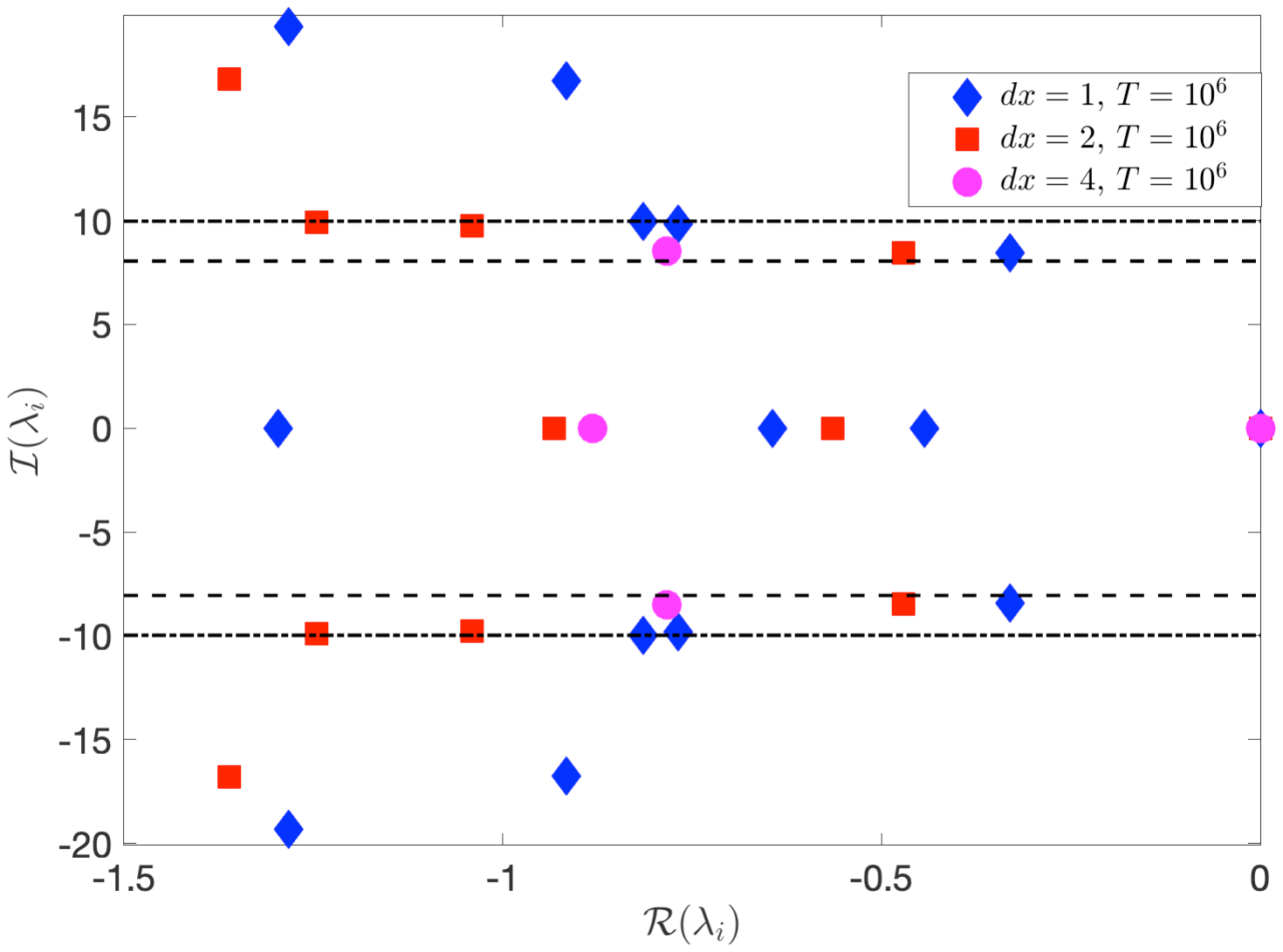}
        \label{fig:spectra_a}
    }
    
    \vspace{0.5cm} 
    
    \subfigure[ Same as (b), but for Vorono\"{i} tessellation resulting from K-means clustering with N centers.]{
        \includegraphics[width=0.48\linewidth]{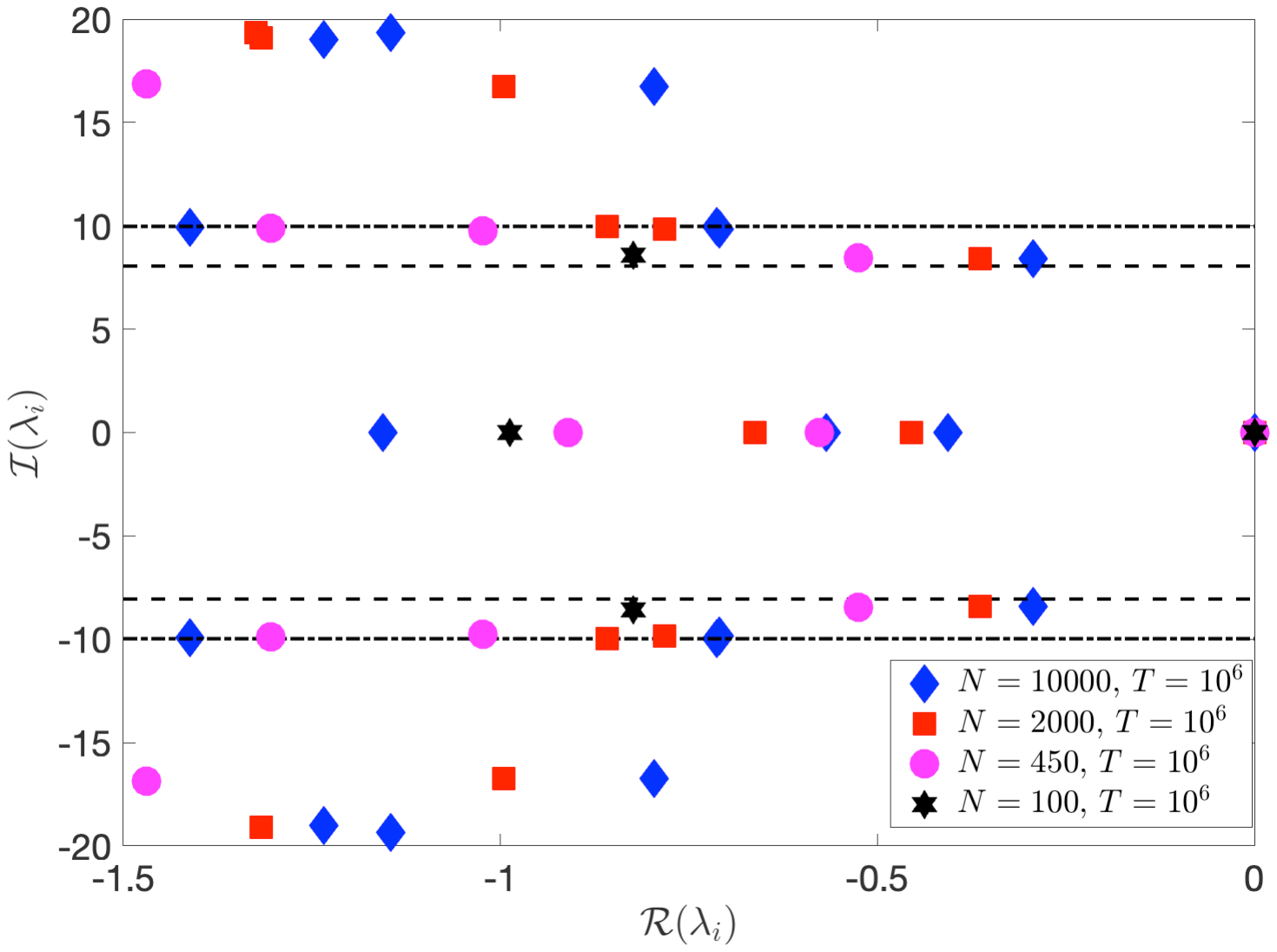}
        \label{fig:spectra_b}
    }
    \hfill
    \subfigure[ Leading $\mathcal{L}_0$ eigenvalues using Chebyshev polynomials dictionaries]{
        \includegraphics[width=0.48\linewidth]{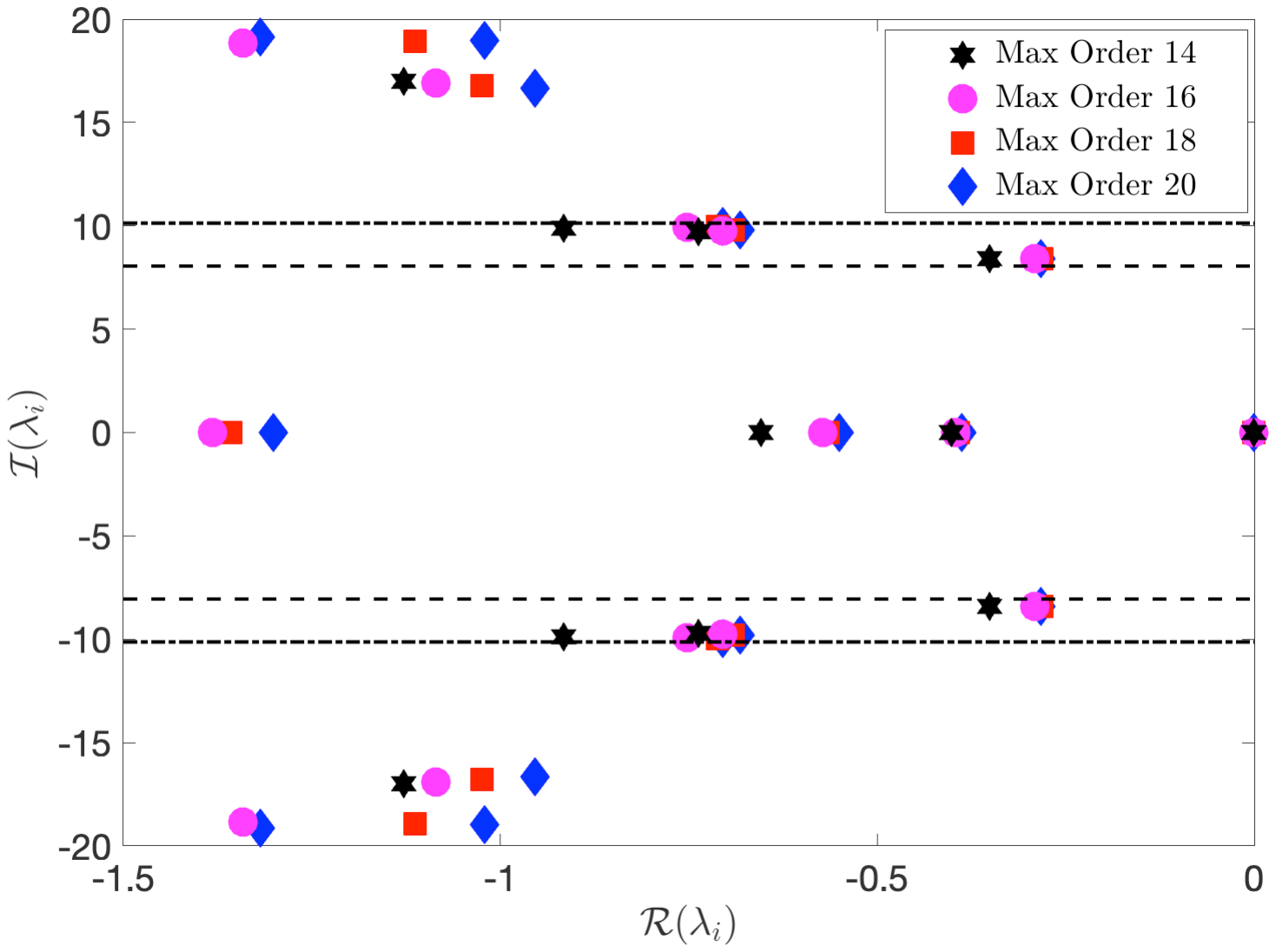}
        \label{fig:spectra_c}
    }
    
    \caption{Magenta, red, and blue dots in panels (b) and (c) indicate results obtained with approximately the same number of cells. 
        The dashed lines in b), c), and d) indicate the values $\pm \pi/p_1$, where $p_1\approx 1.55$ is the period of the shortest UPO of the Lorenz '63 model \cite{Maiocchi2022}. 
        The dot-dashed line indicates the values $\pm 2\pi/q_1$, where $q_1\approx0.62$ is the turnover time around either unstable fixed points apart from the origin \cite{Li2024}. 
        See structure of the corresponding nine leading modes in Fig. \ref{PerronFrobeniusmodes}.}
    \label{spectra}
\end{figure*}

Our data are produced by running the model given in Eq. \ref{Themodel} for  a total of $T=10^6$ time units after disregarding an initial transient with duration of $10^3$ time units. We sample the output with coarse-grained time resolution $d\tau=0.1$ (hence once every 100 time steps), so that we obtain $M=10^7$ points in $\mathbb{R}^3$. The choice of the dictionary is a key step for performing eDMD. We follow here three separate routes in order to showcase pluses and minuses of different approaches:
\begin{enumerate}
    \item Each element of the dictionary $\psi_j=\mathbf{1}_{A_j}$ is the  characteristic function of the set $A_j$, where the $A_j$'s are identically sized cubes of side $\mathrm{d}x$ that partition the phase space, as in \cite{Lucarini2016}. We consider $\mathrm{dx}=1,2,4$. As a result, whilst the support of the invariant measure is the whole of $\mathbb{R}^3$, it turns out that we end up visiting  approximately $N=10000$, $N=2000$ and $N=450$ grid cells, respectively. Note that $N\approx(\mathrm{dx})^{-2}$, similarly to what was observed in \cite{Lucarini2016}, which clearly indicates that within this range of scales, and given the finite sampling, the support of the sampled measure closely resembles its deterministic counterpart. 
    \item Following \cite{Lucarini2025}, each element of the dictionary $\psi_j=\mathbf{1}_{B_j}$ is the  characteristic function of the set $B_j$, where the $B_j$'s are the cells of a Vorono\"i tessellation \cite{Aurenhammer1991}  constructed by performing k-means clustering
\cite{Forgy65,Lloyd82} on the sampled data. Specifically, we use the MATLAB function \texttt{kmeans} using Euclidean distance \cite{MATLAB2024}. By construction, this gridding is data adaptive and the cluster centers are chosen so that the (in this case, Euclidean) distance between the data and the cluster centers is minimized. The metaparameter here is the number of cluster centers $N$. In order to be able to compare the results to case 2., we choose $N=10000$, $N=2000$, and $N=450$, plus a very low-resolution case given by $N=100$.

 \item {Lastly, we follow \cite{Zagli2024} and consider a dictionary of tensorized Chebyshev polynomials of the first kind \cite{Abramowitz1964}. Denoting with $T_n(x)$ the Chebyshev polynomial of the first kind of degree $n$, we consider a dictionary formed by the set $\{ T_i(x)T_j(y)T_k(z) : i + j +k \leq P \}_{i,j,k=1}^P$. We consider $P=14, 16, 18$,  and $20$, which amounts to considering dictionaries comprising $N=680, 969, 1330,$ and $1771$ elements, respectively. The trajectory data is scaled such that all variables are in the interval $[-1,1]$. To obtain reliable spectral properties of the Koopman matrix, we regularize the Gram matrix $\mathbf{G}$ by seeking for a singular value decomposition (SVD) $\mathbf{G = U \Sigma U^T}$ (note that this can be interpreted also as an eigendecomposition because $\mathbf{G}$ is symmetric) and truncating to the first $r$ singular values $\sigma_i = \Sigma_{ii}$ such that $\sigma_i / \sigma_1 \geq \delta$. In the following, we consider $\delta = 10^{-4}$, but we have observed robustness of the results over the interval $\delta \in [10^{-3}, 10^{-8}]$. Employing an SVD corresponds to working in a reduced space spanned by the functions $\hat{\psi}_i = \sum_{j=1}^N U_{ji}\psi_j$ with $i=1,\dots,r$. These functions are constructed to be orthogonal with respect to the (empirical) invariant measure of the system.  The spectral features of the Koopman operator are obtained in this reduced space, with the Koopman eigenfunctions being approximated as a linear combination of the functions $\hat{\psi}_j$ similar to \eqref{eq: eigenfunction}.}

\end{enumerate}
We remark that strategies 1. and 2. amount to analysing the spectral properties of Markov chains associated with the Ulam partitions \cite{Ulam1960,Froyland1997} defined by the space tessellations. Although in the two cases the geometry of the tessellation is very different, one can seamlessly use the formalism developed in \cite{Lucarini2025}. Note also that the statistics of the occupancy $\phi_i=M_i/M$ of the cells, where $M_i$ is the number of times the trajectory is in the $i^{th}$ cell, differ significantly between the two cases, even when partitions with approximately the same total numbers of cells $N$ are chosen. Figure \ref{fig:spectra_d} shows that whilst for the data-adaptive Vorono\"i tessellation all cells have approximately the same population (within an order of magnitude), the rigid structure of the regular cubic gridding leads to very heterogeneous properties in the population of the cells, with values differing by several orders of magnitude.  

We now discuss the estimates for the leading $\lambda_j$'s obtained using the various dictionaries above. Figure \ref{fig:spectra_a} reports the spectra obtained using the dictionary 1. above for various degrees of granularity of the gridding. As we consider finer and finer grids, the absolute value of real part of the $\lambda_j$'s decreases, because the impact of the artificial diffusion associated with the coarse graining \cite{generatorfroyland} is reduced. Despite this, our analysis shows encouraging robustness, because  the first two subdominant eigenvalues are a complex conjugate pair $\lambda_1=\lambda_2^\ast$, followed by two real numbers ($\lambda_3$ and $\lambda_4$), followed by two more pairs of complex conjugate values ($\lambda_5=\lambda_6^\ast$ and $\lambda_7=\lambda_8^\ast$). This ordering is the same regardless of the choice of $\mathrm{dx}$. These features are also seen for dictionaries 2. and 3., as can be seen in Figs. \ref{fig:spectra_b}-\ref{fig:spectra_c}.

\begin{figure*}[t] 
    \centering
    
    \subfigure[Invariant density $\rho_0$.]{
        \includegraphics[width=0.31\linewidth]{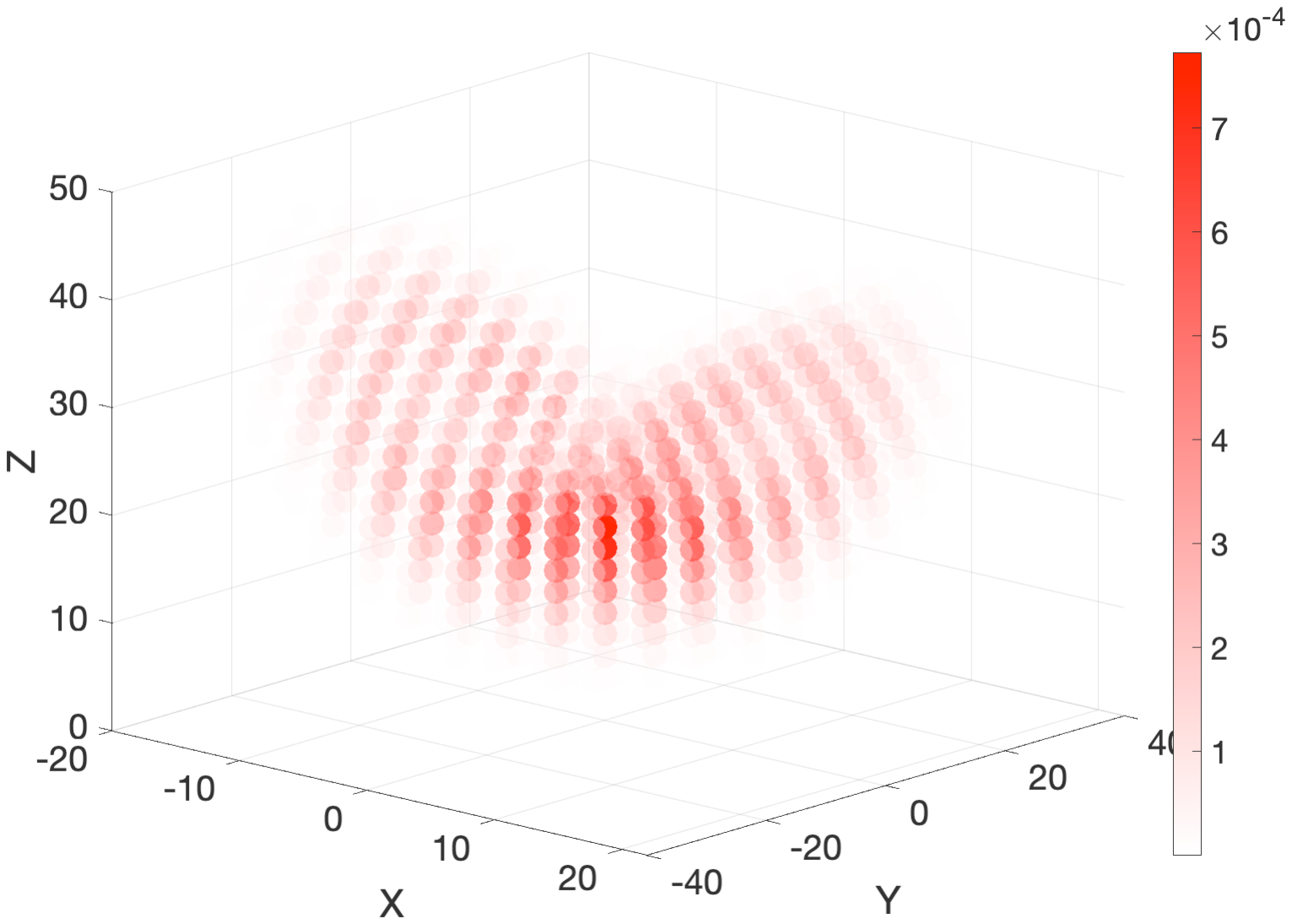}
        \label{fig:rho0}
    }
    \hfill
    \subfigure[ Absolute value of $\rho_1$.]{
        \includegraphics[width=0.31\linewidth]{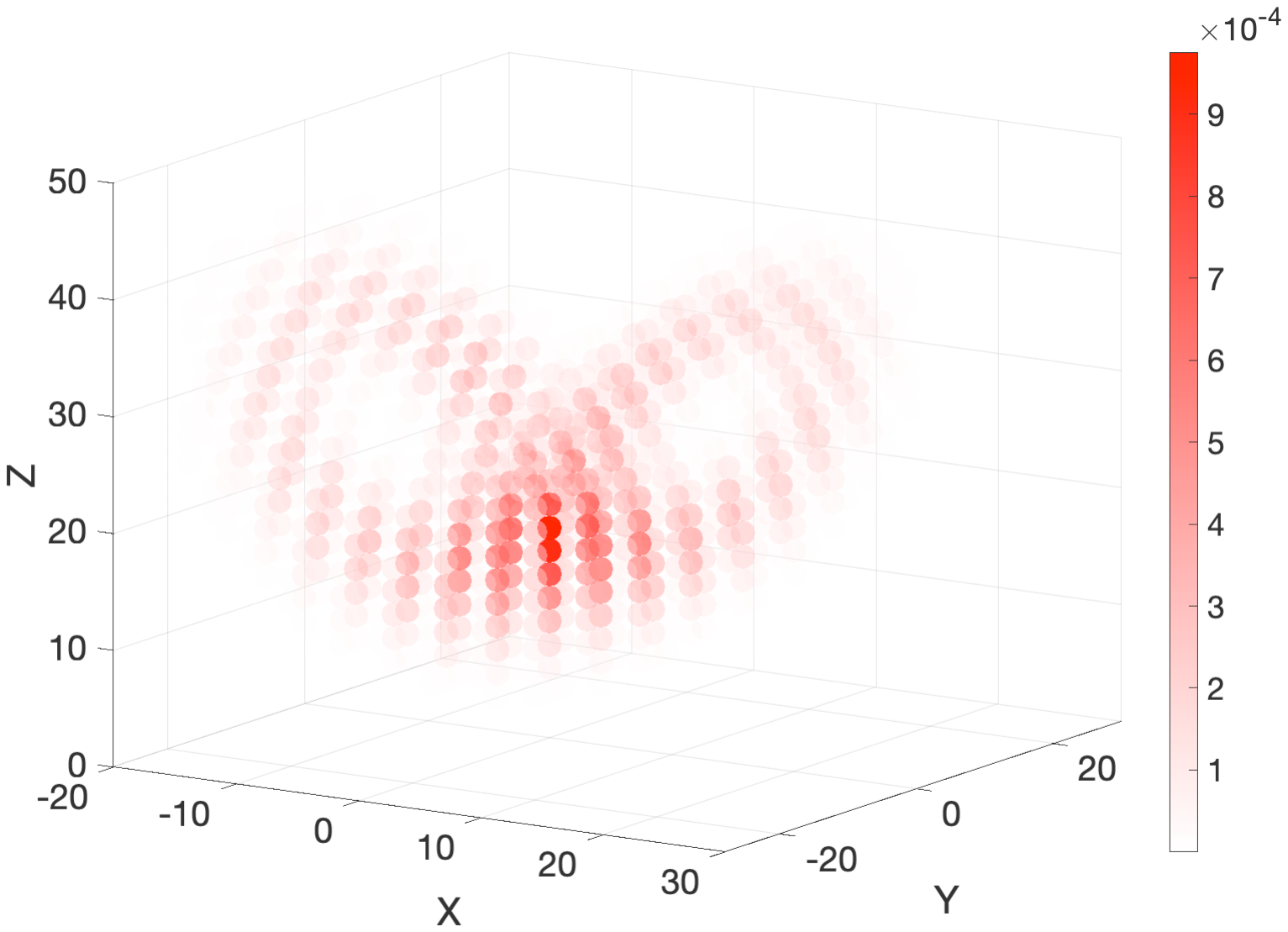}
        \label{fig:absrho1}
    }
    \hfill
    \subfigure[ Phase of $\rho_1$.]{
        \includegraphics[width=0.31\linewidth]{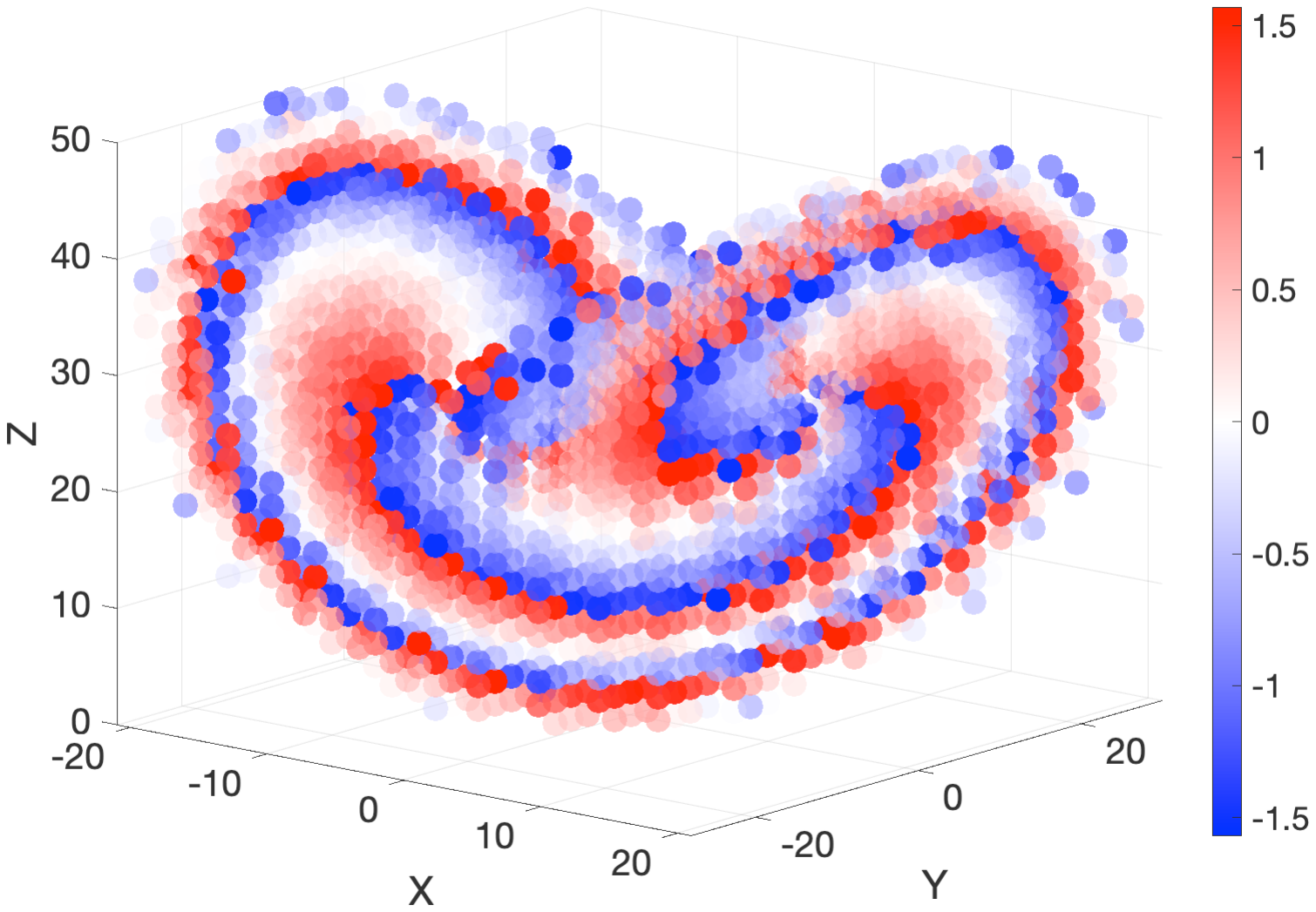}
        \label{fig:phrho1}
    }
    
    \vspace{0.3cm} 
    
    \subfigure[ $\rho_3$.]{
        \includegraphics[width=0.31\linewidth]{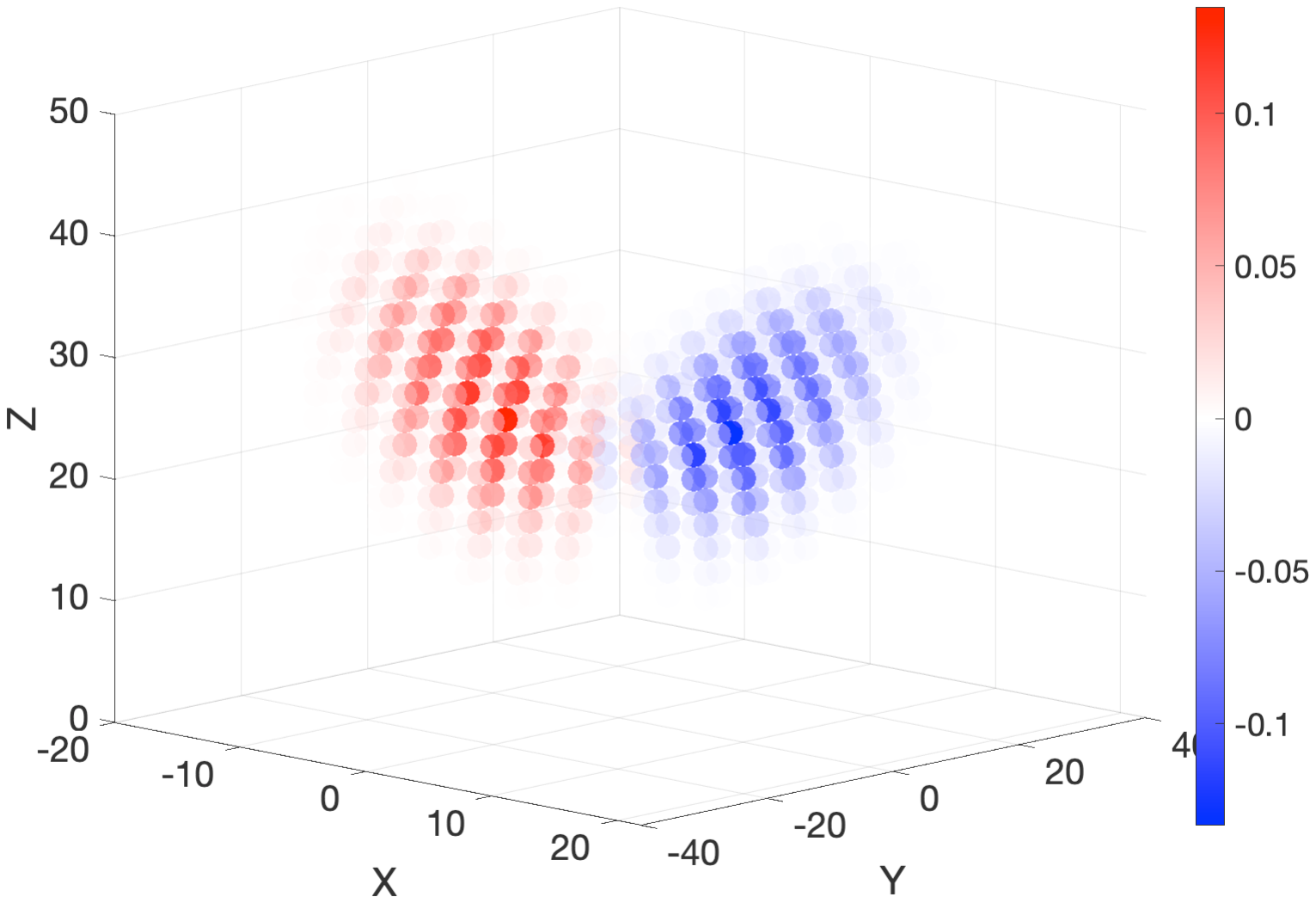}
        \label{fig:rho3}
    }
    \hfill
    \subfigure[ $\rho_4$.]{
        \includegraphics[width=0.31\linewidth]{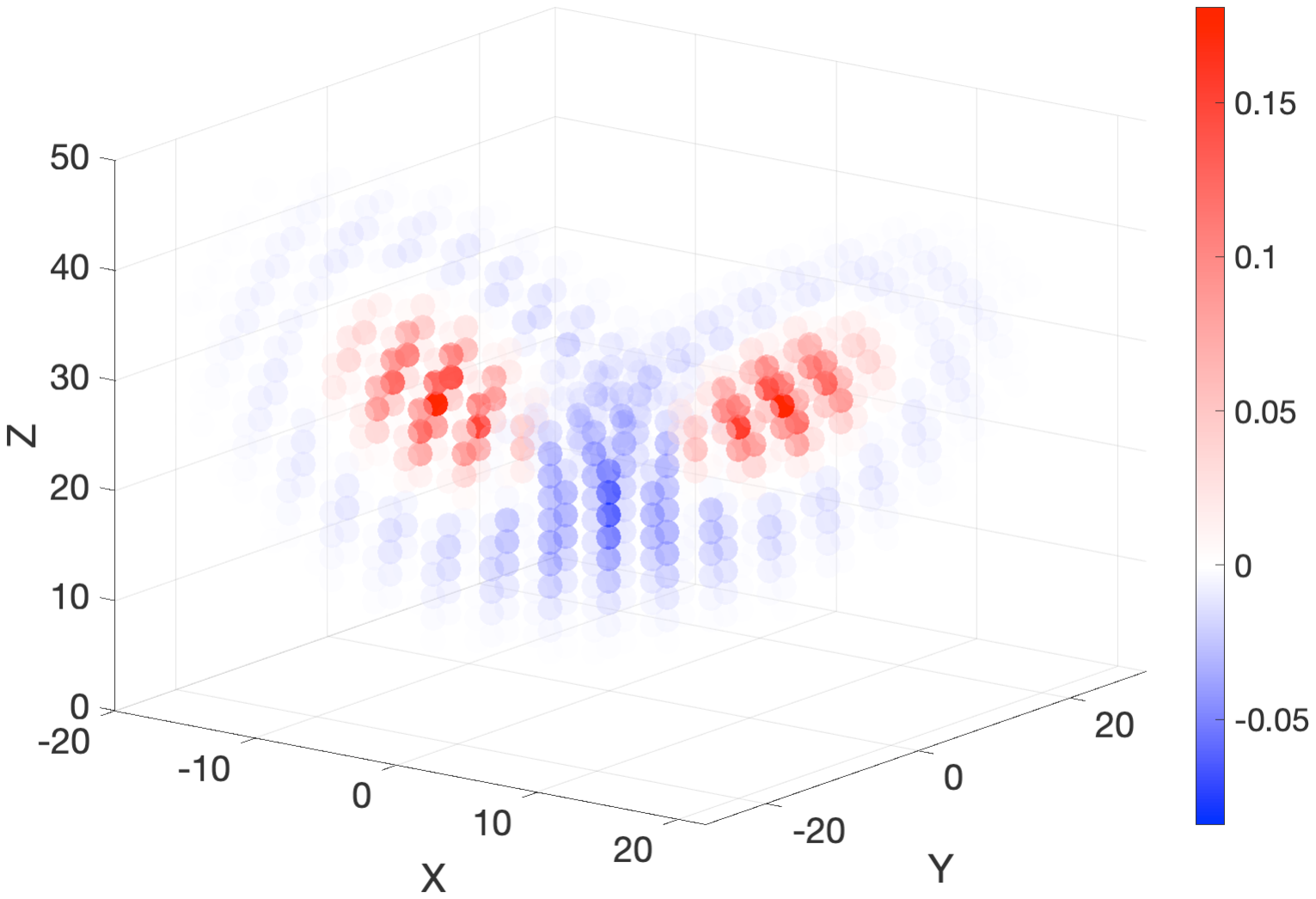}
        \label{fig:rho4}
    }
    \hfill
    \subfigure[ Absolute value of $\rho_5$.]{
        \includegraphics[width=0.31\linewidth]{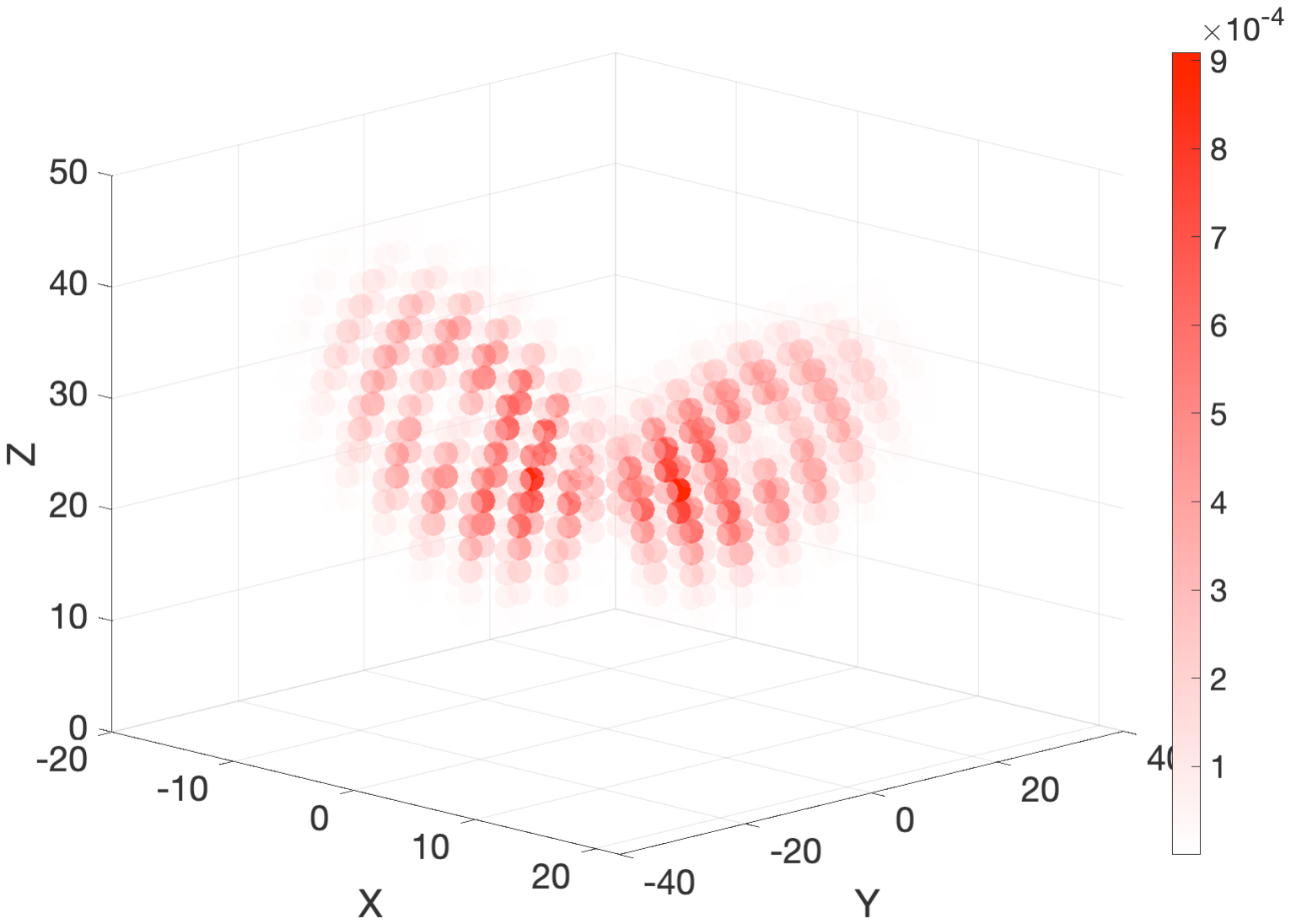}
        \label{fig:absrho5}
    }
    
    \vspace{0.3cm} 
    
    \subfigure[ Phase of $\rho_5$.]{
        \includegraphics[width=0.31\linewidth]{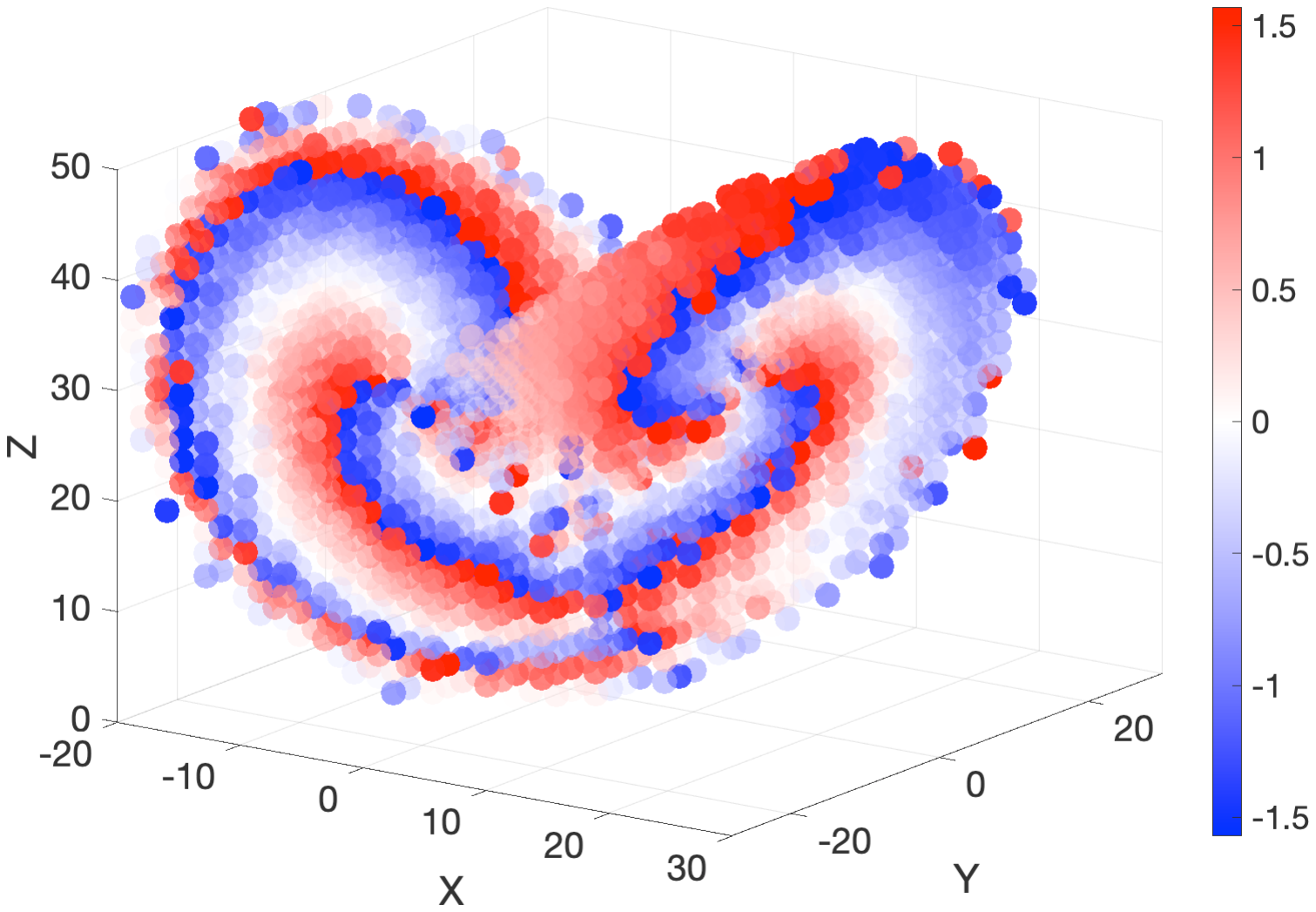}
        \label{fig:phrho5}
    }
    \hfill
    \subfigure[ Absolute value of $\rho_7$.]{
        \includegraphics[width=0.31\linewidth]{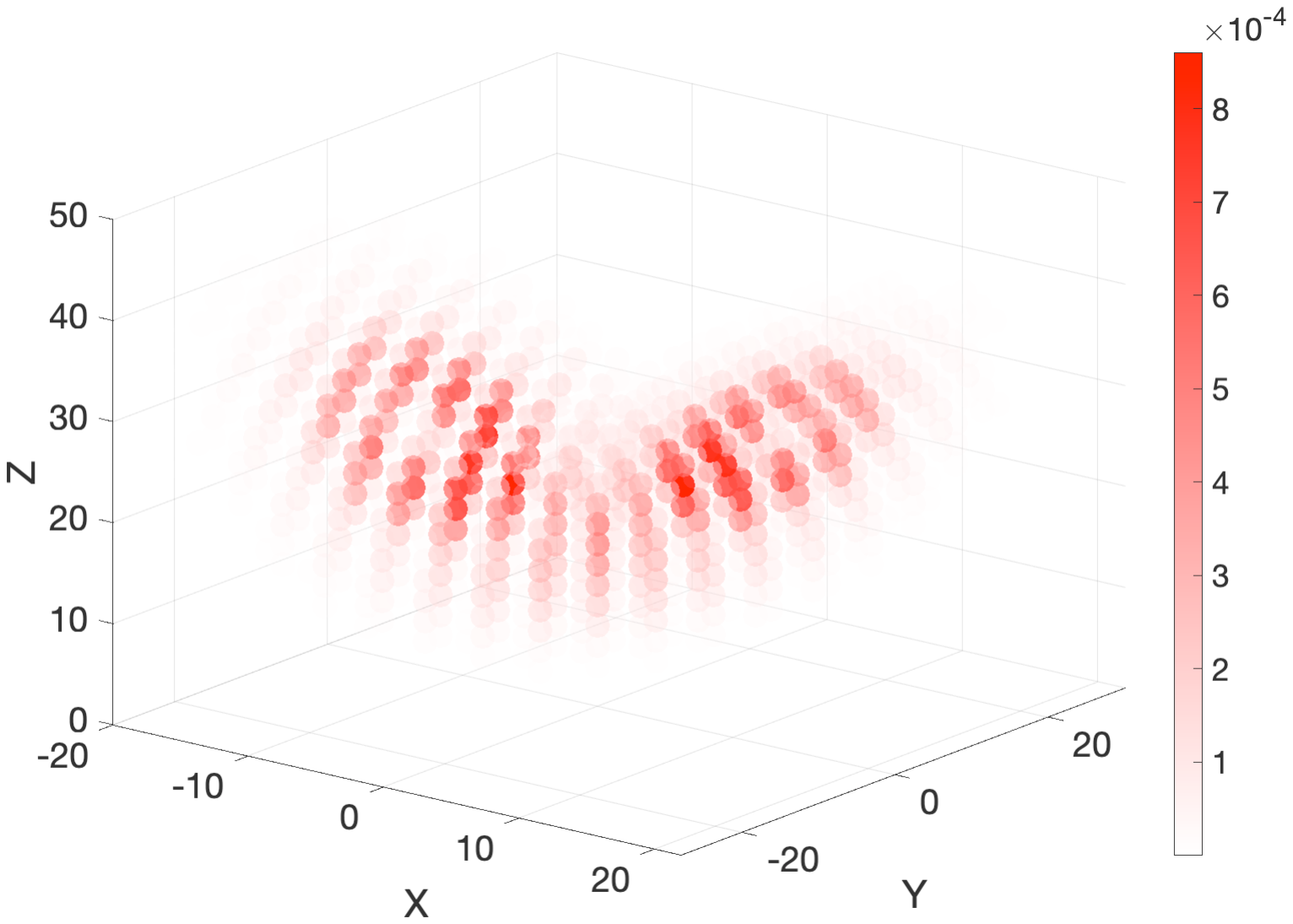}
        \label{fig:absrho7}
    }
    \hfill
    \subfigure[ Phase of $\rho_7$.]{
        \includegraphics[width=0.31\linewidth]{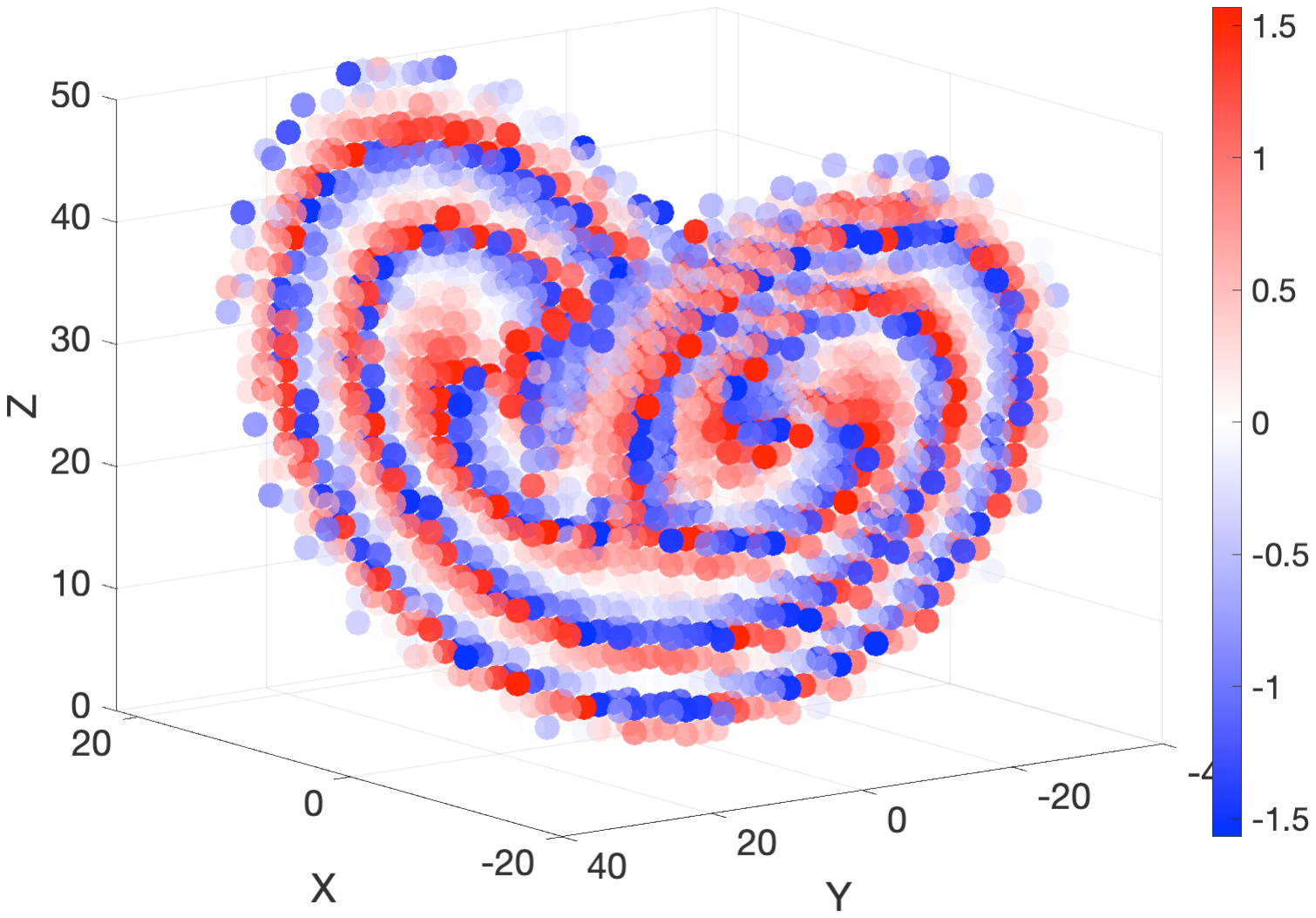}
        \label{fig:phrho7}
    }
    
    \caption{Nine leading modes of the Perron-Frobenius operator obtained using the regular Ulam partition with $\mathrm{dx}=2$, see Fig. \ref{fig:spectra_a} for the corresponding eigenvalues.}
    \label{PerronFrobeniusmodes}
\end{figure*}

We also note that  $\rho_3$ and $\rho_4$, shown in Figs. \ref{fig:rho3} and \ref{fig:rho4} describe the redistribution of mass between the regions associated with the (weakly) unstable fixed points of the determistic Lorenz '63 system (which become stable for $r\approx23.5$ \cite{Tantet2018}) and between the inner and outer regions of the two wings of the butterfly, respectively.

Additionally, we have a very robust estimate of the imaginary part of the $\lambda_j$'s for the complex conjugate pairs. For all considered values of $\mathrm{dx}$, $\mathcal{I}(\lambda_5)=\mathcal{I}(\lambda_7)\approx 2\pi/q_1$, where $q_1\approx0.62$ is the turnover time around either unstable fixed points at the center of the wings of the butterfly \cite{Li2024}. Instead, $\mathcal{I}(\lambda_1)\approx \pi/p_1$  where $p_1\approx1.55$ is the period of the shortest UPO of the Lorenz 63 model \cite{Maiocchi2022}. Indeed, in this case what matters is half of the UPO trajectory, associated with the twirling behaviour on either side of the attractor. Instead, closed UPOs are global objects for the Lorenz '63 system, which link the two sides of the attractor, hence contributing directly to the transfer of measure mainly associated with the real component of the Koopman spectrum \cite{Maiocchi2022}.

Note that the imaginary parts of other $\lambda_j$ reported in Fig. \ref{fig:spectra_a} are twice the values reported above, thus indicating higher harmonic contributions. This shows that the leading complex modes describing decay of correlation are closely related to the rotational features of the deterministic dynamics. 

This can be further clarified  when we observe the spatial patterns of the corresponding modes of the Perron-Frobenius operator. Keep in mind that if $\lambda_j=\lambda_k^\ast$, then $\rho_j=\rho_k^\ast$. The properties of the modes portrayed in Fig. \ref{PerronFrobeniusmodes} are robust across the considered values of $\mathrm{dx}$. 

The phases of  $\rho_1$, $\rho_5$, and $\rho_7$ shown in Figs. \ref{fig:phrho1}, \ref{fig:phrho5}, and \ref{fig:phrho7}, respectively, clearly show the rotational component of the dynamics (which is an unavoidable signature of nonequilibrium). We can capture, however, the key difference between $\rho_1$, on one side, and  $\rho_5$, and $\rho_7$, on the other side, as follows. Whilst $|\rho_1|$, shown in Fig. \ref{fig:absrho1}, peaks in the region between the two wings of the butterfly, hence showing that its associated with motions connecting the two distinctive sides of the attractor,    $|\rho_5|$ and $|\rho_7|$, which are shown in Figs. \ref{fig:absrho5} and \ref{fig:absrho7}, have larger values in the regions immediately surrounding either eye of the butterfly, thus indicating relevance for a more localized dynamics.

The features described above are coherently found  when considering the dictionary 2, see Fig. \ref{fig:spectra_b}. In this case, we have a faster convergence of the value of the real part of the eigenvalues $\lambda_j$ as the partition becomes more and more refined. In other terms, the data-adaptive dictionary seems to perform  better than the one associated with a regular cubic tessellation, as the effect of the artificial diffusion is greatly reduced. 

The estimates of the eigenvalues we obtain using {the Chebyshev polynomials} dictionary - see Fig. \ref{fig:spectra_c} - seem even more robust with respect to changes in the maximum order of the monomials, with a lower number of dictionary elements needed to achieve convergence compared to the case of Ulam partitions.

\section{Response to Perturbations}\label{responseoperators}
We consider a standard perturbation of the drift term of the form $\mathbf{F}(\mathbf{x})\rightarrow \mathbf{F}(\mathbf{x})+\varepsilon \mathbf{X}(\mathbf{x})h(t)$ where $\mathbf{X}(\mathbf{x})=(0,x,0)$. The function $h(t)$ is a bounded function representing the time-modulation of the forcing $\mathbf{X}$. The applied perturbation  amounts to a change $r\rightarrow r+\varepsilon h(t)$ in the model given in Eq. \ref{model}. The Fokker-Planck equation \eqref{FPE} now reads as follows:
\begin{equation}
\partial_t\rho_\varepsilon(\mathbf{x},t)=\mathcal{L}_0\rho_\varepsilon(\mathbf{x},t){+\varepsilon\mathcal{L}_1\rho_\varepsilon(\mathbf{x},t) h(t)}
\end{equation}
where $\mathcal{L}_1 \bullet =-\nabla \cdot(\mathbf{X}(\mathbf{x}) \bullet)$. We shall assume that the perturbation operator $\mathcal{L}_1$ shares the same domain as $\mathcal{L}_0$, and that $\mathcal{L}_0 + \varepsilon \mathcal{L}_1$ is the infinitesimal generator of a $C_0$-semigroup. 
The task of response theory is to express expectation values of observables with respect to $\rho_\varepsilon$ in terms of $\rho_0$. Recall that, in practice, $\rho_\varepsilon$ is only accessible by solving the perturbed dynamics; however linear response avoids this costly step. Denoting $\left\langle f \right\rangle_\varepsilon=\int\mathrm{d}\mathbf{x}f(\mathbf{x})\rho_\varepsilon(\mathbf{x},t)$ as the expectation value of $f$ with respect to $\rho_\varepsilon$, we have the following linear response relation
\begin{equation}\label{LRF}
    \frac{\mathrm{d}\langle f \rangle_\varepsilon}{\mathrm{d}\varepsilon}|_{\varepsilon=0}=\varepsilon  (h \star G_{f,\mathbf{X}}) (t).
\end{equation}
where ``$\star$" denotes the convolution and where \begin{equation}\label{Greengeneral}
    G_{f,\mathbf{X}}=\Theta(t)\int  \mathrm{d}\mathbf{x} \mathcal{L}_1\rho_0(\mathbf{x}) e^{t\mathcal{L}_0^{\ast}}f(\mathbf{x})= \Theta(t)\int  \mathrm{d}\mathbf{x}\rho_0(\mathbf{x})\Gamma(\mathbf{x})  e^{t\mathcal{L}_0^{\ast}}f(\mathbf{x})
\end{equation} 
is the so-called Green's function, which can be interpreted as $t-$lagged the correlation between the observable $f(\mathbf{x})$ and the observable $\Gamma(\mathbf{x})=\mathcal{L}_1\rho_0(\mathbf{x})/\rho_0(\mathbf{x})$ in the unperturbed state for positive time lags  \cite{Marconi2008,Santos2022}. Causality is enforced by the Heaviside distribution $\Theta(t)$.
The Green's function describes, apart from the factor $\varepsilon$, the response of the system to a Dirac $\delta$ impulse at $t=0$ and its decay properties reveal the sensitivity of the system to the applied field $\mathbf{X}$. The convolution with $h(t)$, as shown in Eq.~\eqref{LRF}, gives the leading-order correction of $\left\langle f \right\rangle_\varepsilon$ in powers of $\varepsilon$. 

We consider the perturbation to $r$ discussed above and, following \cite{Reick2002,Lucarini2009DispersionNonLinearResponseLorenz,Lucarini2016}, analyze for simplicity the observable $f(\mathbf{x})=z$. Indeed, $x^2$, $y^2$, $z^2$ and $z$ play a very special role for determining the dynamical properties of the Lorenz '63 system and its energetics \cite{Blender2013,Pelino2014}. The Green's function of interest reads as 
\begin{equation}\label{Greenehere}
    G_{z,r}(t)=\Theta(t)\int \mathrm{d}\mathbf{x}  \mathcal{L}_1\rho_0(\mathbf{x}) e^{\mathcal{L}_0^{\ast}t}z=-\Theta(t)\int \mathrm{d}\mathbf{x}\rho_0(\mathbf{x})x \partial_y \log(\rho_0(\mathbf{x}) )e^{\mathcal{L}_0^\ast t} z,
\end{equation}
where we have applied the definition of $\mathcal{L}_1$.

The existence of a linear response relation amounts to determining whether the convolution in Eq.~\eqref{LRF} converges. This is guaranteed if, as in our case, the operator $\mathcal{L}_0^{\ast}$ has a spectral gap.  
As a result, we {obtain \cite{Santos2022}:
\begin{equation}\label{Greenalpha}
 G_{f,X}(t)=\Theta(t)\sum_{j=1}^\infty \alpha_j\exp(\lambda_j t) \quad \alpha_j = \int \mathrm{d}\mathbf{x}\rho_j(\mathbf{x})\varphi_j(\mathbf{\mathbf{x}})f(\mathbf{x}) \mathcal{L}_1\rho_0(\mathbf{x}),  \end{equation}}
The Green's function features exponential decay if we have a non-vanishing spectral gap, i.e. $\mathcal{R}(\lambda_1)<0$. Note also that the $j=0$ term in the summation can be shown to cancel out \cite{Santos2022}. Additionally, the static sensitivity of the system to perturbation, which is given by Eq. \ref{LRF} when one considers $\lim_{t\rightarrow\infty}$ and $h(t)=1$, can be written explicitly as $\mathrm{d}\langle f\rangle _\varepsilon/\mathrm{d}\varepsilon|_{\varepsilon=0} =\int_0^\infty \mathrm{d}t G_{f,X}(t)=\sum_{j=1}^\infty\alpha_j/\lambda_j$, which underlines that the slowest decaying modes play a key role in determining the sensitivity of the system; see also the detailed calculations in \cite[Chap. 2]{santos2022phd}. Conversely, 
the fastest decaying modes can play an important role at initial times. In the limit of $t$ tending to zero from the right we get the following expression for the derivatives of Green function
\begin{equation}\label{short}
     \frac{\mathrm{d}^p G_{f,X}(t)}{\mathrm{d} t^p}|_{t=0^+}=\sum_{j=1}^\infty \alpha_j \lambda_j^p,
\end{equation}    
which shows that     
the relevance of the rapidly decaying part of the spectrum becomes apparent as we consider higher and higher derivatives. 
Notice that geometric multiplicities  of the eigenvalues (omitted here) might result in extended transients before the exponential decay takes over. 

\subsection{Computing the Green's Function}
\subsubsection{Direct Numerical Simulation}
In order to compute the Green's function given in Eq. \ref{Greenehere}, we follow a variety of methods. First, we use the definition given in Eq. \ref{LRF}. We run a long simulation of the unperturbed model lasting $2\times10^7$ time units, from which a) we evaluate the expectation value of $z$ and b) we extract $10^6$  temporally equispaced sample points. The time separation between such sample points is much larger than  $1/\mathcal{R}(\lambda_1)$, see Fig. \ref{spectra}. Hence, these samples provide good candidates from (approximately) independent initial conditions that sample the invariant measure. We then apply to each initial condition the perturbations $\mathbf{X}(\mathbf{x})h(t)$ where $h(t)=\varepsilon/2 \delta(t)$, with $\varepsilon=0.1$, and integrate the system for 20 time units. We perform a second set of integrations as above where this time $\varepsilon=-0.1$. We take the difference between the value of $z(t)$ in each pair of integrations, average over the $10^6$ samples, and divide such an average by $\varepsilon$. Using centred differences greatly improves the estimate of the linear response terms, as as quadratic corrections are eliminated \cite{gritsun2017}. The obtained estimate of $G_{z,r}$ is shown in Fig. \ref{fig:sub4}. With some extra runs, we  have tested that the chosen value of $\varepsilon$ is well within the range of linearity of the response, so that what is portrayed in Fig. \ref{fig:sub4} is our reference result. 

Next we use the definition given in Eq. \ref{Greenalpha}, hence using the knowledge gathered via Koopman analysis on the unperturbed system to predict its response to perturbations. 

\subsubsection{Equation-Free, Data-Driven Approach}

As a first option, we do not want to  explicitly use our knowledge of the reference governing equations of the system as well as of the perturbation vector field for the goal of constructing the response operator. Hence, we target a data-driven approach to the problem. We consider two additional steady-state model runs performed according to the same protocol described in Sect. \ref{koopman} but using $r=28.5$ and $r=27.5$ (so that $\delta r=1$) respectively. We then apply in each case the Koopman analysis discussed in section \ref{koopman} using dictionary 1. to derive the transfer operator acting for a time $\mathrm{d}\tau$. We refer to such operators as $\Pi^{\mathrm{d}\tau}_+$ (for $r=28.5$) and $\Pi^{\mathrm{d}\tau}_-$ (for $r=27.5$), and define $\Delta\Pi_0^{\mathrm{d}\tau}= \Pi^{\mathrm{d}\tau}_+-\Pi^{\mathrm{d}\tau}_-$. Since we are now considering a discretized state space, following \cite{Lucarini2025}, we define $\mathcal{M}=\Pi^{\mathrm{d}\tau}_0\in\mathbb{R}^{N\times N}$ as the stochastic matrix describing the evolution of the Markov chain defined over the states $A_j,$ $j=1,\ldots,N$ over a time interval $\mathrm{d}\tau$, and $m=\Delta\Pi_0^{\mathrm{d}\tau}\in\mathbb{R}^{N\times N}$ as its perturbation. Note also that in this cased we have used centred differences in order to improve the quality of our estimate of the linear perturbation operator and as a result $\delta r=1$ is well within the range of linearity. Substituting the definitions above into  Eq. \ref{Greengeneral} we obtain the following formula: 
\begin{equation}
    G_{f,m}(q\mathrm{d}\tau)=(\mathrm{d}\tau)^{-1}\Theta(t)\langle \left(\mathcal{M}^T\right)^q f,m \rho_0\rangle=(\mathrm{d}\tau)^{-1}\Theta(t)\langle \left(\mathcal{M}^T\right)^q f, m\rho_0\rangle
    \end{equation}
where  $f,\rho_0\in\mathbb{R}^{N}$, the perturbation field $X$ is implicitly represented by $m$, and $\langle \bullet,\bullet \rangle$ indicates the standard scalar product in finite dimensional Euclidean vector space. We can further expand the formula along the lines of Eq. \ref{Greenalpha} with $\alpha_j=\langle   v_j w_j^T f,m\rho_0\rangle$
where $v_j$ and $w_j$ are the $j^{th}$ left and $j^{th}$ right eigenvector of $\mathcal{M}$, respectively, both corresponding to the eigenvalue $\lambda_j$ \cite{Lucarini2025}. The results of this procedure are reported in Fig. \ref{fig:sub1} for different values of the regular gridding $\mathrm{d}x$. Comparing this with Fig. \ref{fig:sub4}, one observes that the oscillatory behaviour is extremely well reproduced, both in terms of frequency and phase, which points to an accurate representation of the imaginary part of the leading eigenvalues, as discussed in Sect. \ref{koopman}. Instead, the decay rate is overestimated, as a result of the artificial diffusion associated with the procedure of coarse graining \cite{generatorfroyland}. Such a bias is indeed reduced as one considers lower values of $\mathrm{d}x$, with the $\mathrm{d}x=1$ case featuring a good performance.

Whilst the long-time behaviour of the Green's function is determined by the first subdominant pair of eigenvalues, the short-time  term behavior is influenced by eigenvalues that lie deeper in the spectrum - see the inset in Fig. \ref{fig:sub1} where we show how the estimate of the Green's function changes as we cut off the summation in Eq. \ref{Greenalpha} at different values $j_{max}$. In such a regime, the value of $\mathrm{d}x$ is a minor factor (not shown). 

The methodology above is repeated step-by-step in an identical fashion in the case of dictionary 2. We find that the use of a data-adaptive partition of the phase space leads to improved agreement with the direct estimate of the Green's function, see Fig. \ref{fig:sub2}. This is mainly due a slower decay of the Green's function compared to the previous case, in agreement with what is observed in Fig. \ref{spectra} regarding the properties of the leading $\lambda_j$'s, with convergence apparently achieved for lower values of $N$. The higher quality of the results obtained using  Vorono\"i tessellations associated with k-means clustering is indeed encouraging in terms of equation-agnostic approaches to the problem.

\subsubsection{Koopmanism-informed Fluctuation-Dissipation Relation}

Finally, we consider dictionary 3. above, which is composed of differentiable functions, in contrast with the two previous dictionaries. Here, we take advantage of our knowledge of the applied perturbation, and hence of the functional form of the operator $ \mathcal{L}_1$. In this case we do not need to observe a run of the perturbed dynamics, but we will make solely use of the Koopman features of the unperturbed dynamics and use exclusively the dataset produced when constructing the Koopman spectrum shown in Sect. \ref{koopman}. 

To obtain the coefficients $\alpha_j$, and, in particular, the decomposition of the response observable $\Gamma = \mathcal{L}_1 \rho_0 / \rho_0$ onto the Koopman eigenfunctions we follow Method 3. introduced in \cite{Zagli2024}. Firstly, we seek a decomposition of $\Gamma(\mathbf{x})$ onto the dictionary functions by minimising the $L^2_{\rho_0} $ distance between $\Gamma(\mathbf{x})$ and the subspace spanned by the dictionary functions, i.e., $\int \mathrm | \Gamma(\mathbf{x}) - \sum_{k=1}^N \Gamma_{k} \psi_k(\mathbf{x}) |^2 \rho_0(\mathbf{x}) \mathrm{d}\mathbf{x}$. This results into a vector of coefficients $\boldsymbol{\Gamma} = \mathbf{G}^{^+}\boldsymbol{\Delta} $ where $\Delta_i = \int\mathrm{d}\mathbf{x}\rho_0(\mathbf{x})\Gamma(\mathbf{x})\psi_i^\ast(\mathbf{x})=\int\mathrm{d}\mathbf{x}\rho_0(\mathbf{x})\mathbf{X}(\mathbf{x})\cdot \nabla \psi_i^\ast(\mathbf{x})$. Using ergodicity, this projection can be estimated by evaluating the action of the (adjoint of) the perturbation operator along a long trajectory \cite{Zagli2024}; specifically, one can estimate it as $ \Delta_i \approx \frac{1}{T}\int_0^T \mathbf{X}(\mathbf{x}(t)) \cdot \nabla \psi_i^*(\mathbf{x}(t)) \mathrm{d}t = \frac{1}{T}\int_0^T x(t)\partial_y\psi_i (t) \mathrm{d}t$ where the last equality holds because the dictionary of functions is real and we have used the explicit form of the perturbation field. 


Since the dictionary of Chebyshev polynomials is closed under the differentiation operator, we can write $\partial_y \psi_i = \mathbf{D}\psi_i$, where $\mathbf{D}$ is a suitable matrix \cite{Abramowitz1964}.  The observable $\Gamma(\mathbf{x})$ is first projected onto the space spanned by the SVD functions $\{ \hat{\psi}_j\}_{j=1}^r$ described in section \ref{koopman}, and then onto the Koopman eigenfunctions. As opposed to the previous response reconstruction methodologies where the left eigenvectors of the Koopman matrix were used to evaluate the coefficients $\alpha_j$, we here perform a projection of $\Gamma(\mathbf{x})$ in the Koopman space by minimising, similarly to above, the $L^2_{\rho_0}$ distance between $\Gamma$ and the space spanned by the first $n$ Koopman modes. 
This projection operation makes sole use of the first $n$ Koopman eigenfunctions $\varphi_n(\mathbf{x})$. In particular, it avoids using the left eigenvectors of the Koopman matrix which depend on all modes, including the deep, poorly resolved ones. By performing this projection operation onto the leading Koopman eigenfuctions, we can safely remove the effect of the deep ones.

\begin{figure*}[ht!]
    \centering
    \subfigure[Ulam Method for regular gridding with cubes of size $\mathrm{dx}$ for $f(\mathbf{x})=z$. Inset shows the impact of the truncation level $M$ in the Koopman expansion for the $\mathrm{dx}=2$ case.]{
        \includegraphics[width=0.48\linewidth]{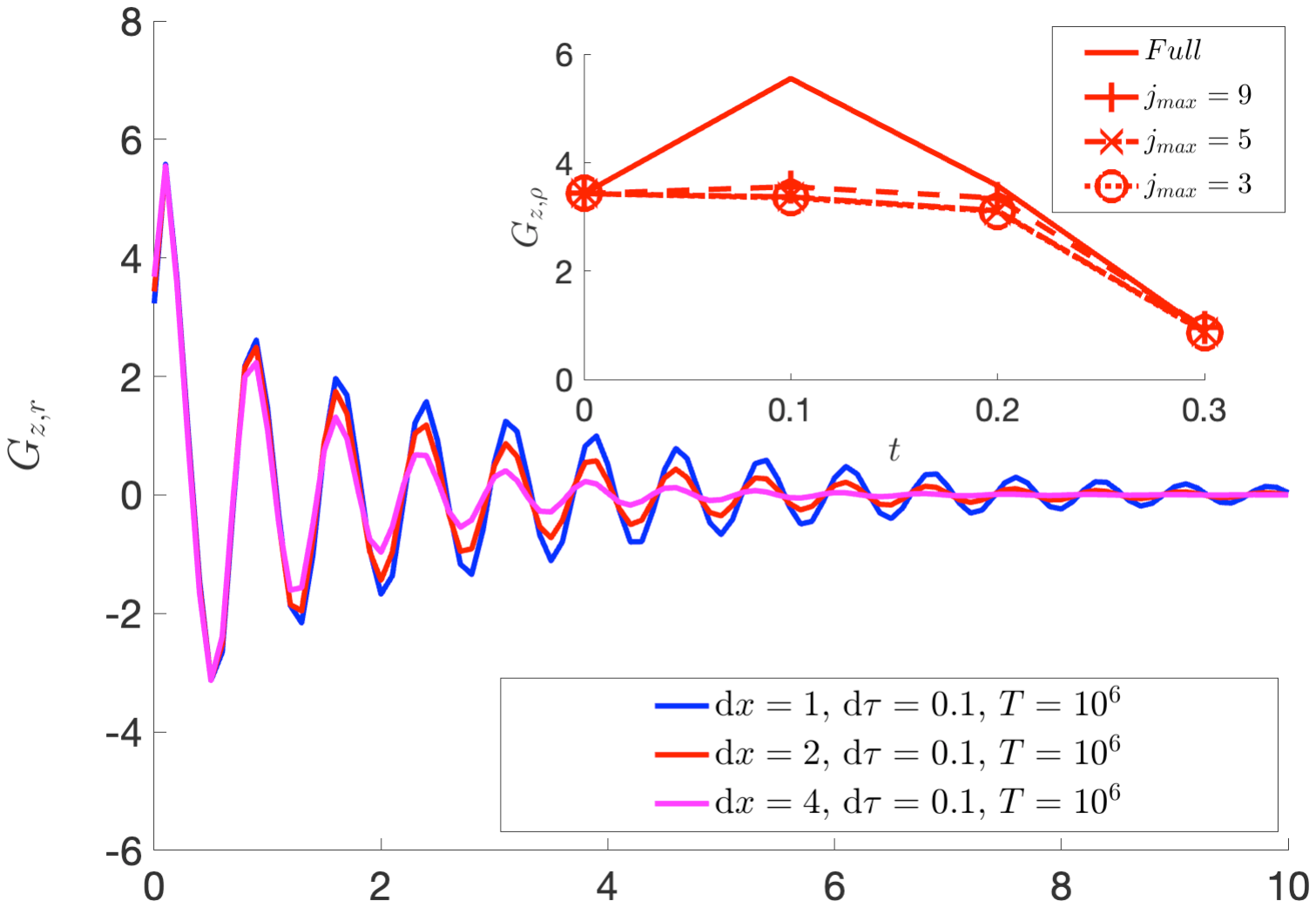}
        \label{fig:sub1}
    }
    \hfill
    \subfigure[Same as (a), but partition made with K-means clustering with different numbers of clusters $N$.]{
        \includegraphics[width=0.48\linewidth]{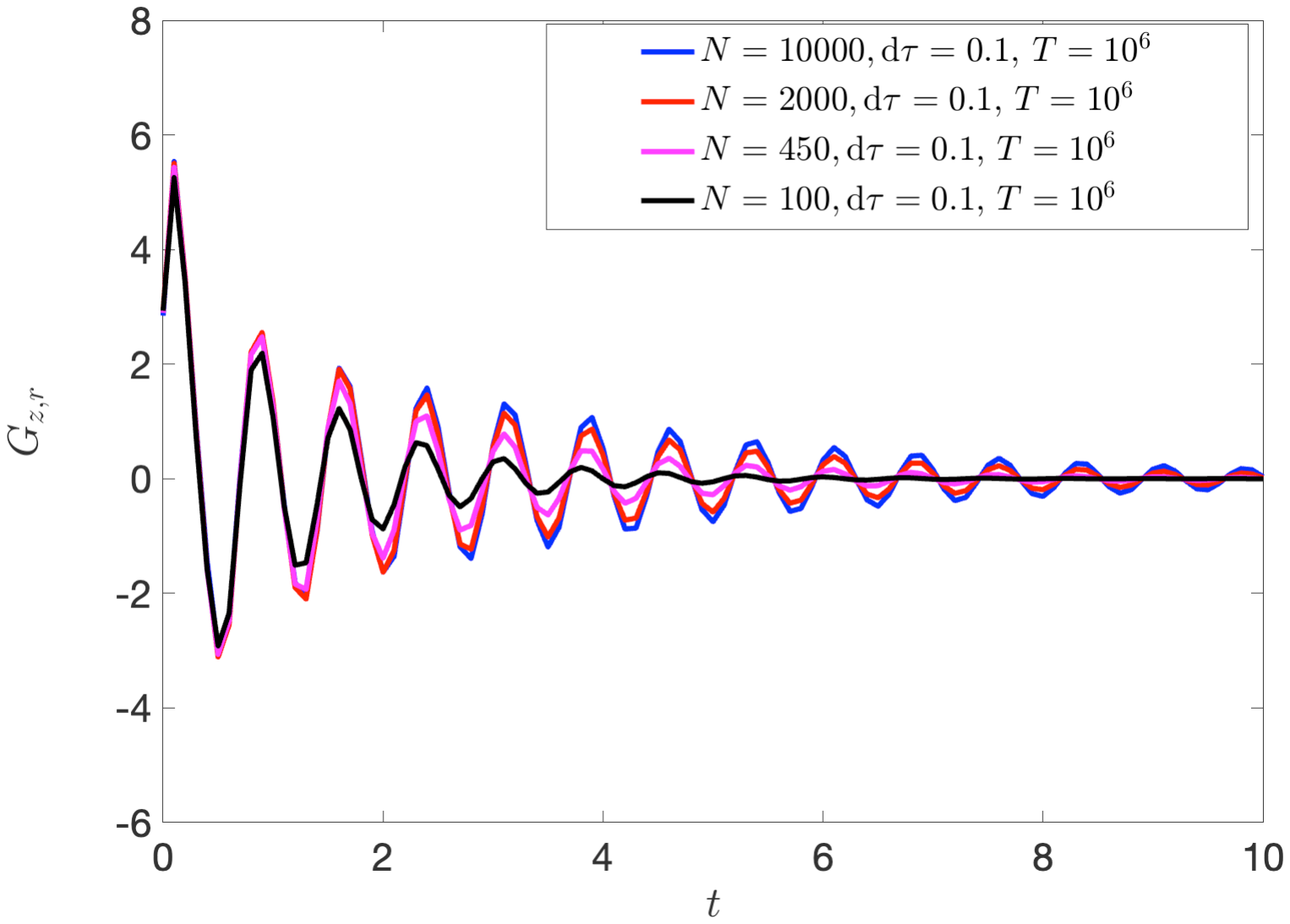}
        \label{fig:sub2}
    }
    
    \vspace{0.3cm}
    
    \subfigure[Reconstruction obtained using as Koopman dictionary Chebyshev polynomials of  maximum order $P$. Only modes with $\mathcal{R}(\lambda_j)>co=-1.5$ are used. Inset shows the impact of the choice of the cutoff $co$, $P=18$ case.]{
        \includegraphics[width=0.48\linewidth]{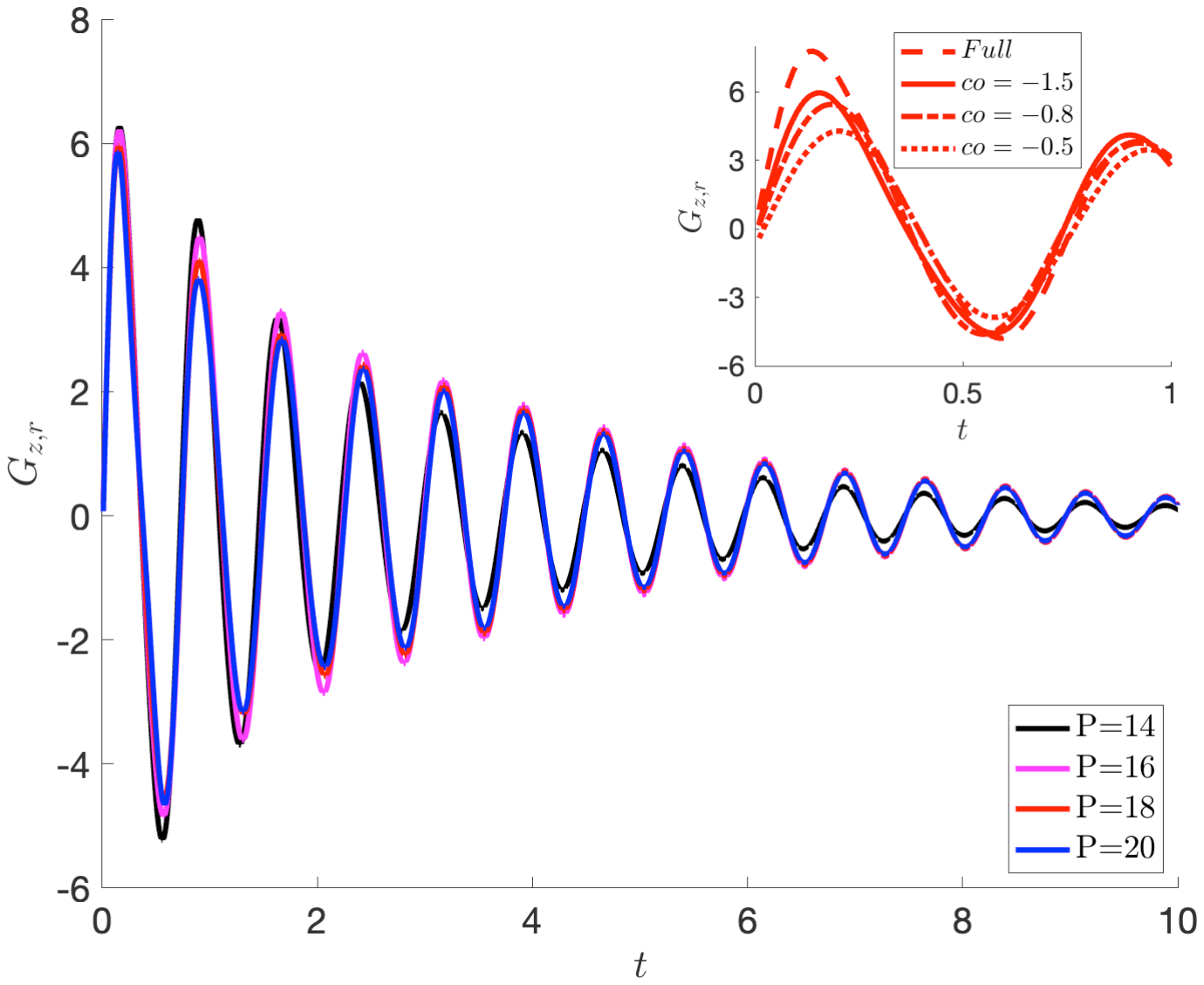}
        \label{fig:sub3}
    }
    \hfill
    \subfigure[Results from direct numerical simulations, average over $10^6$ ensemble members.]{
        \includegraphics[width=0.48\linewidth]{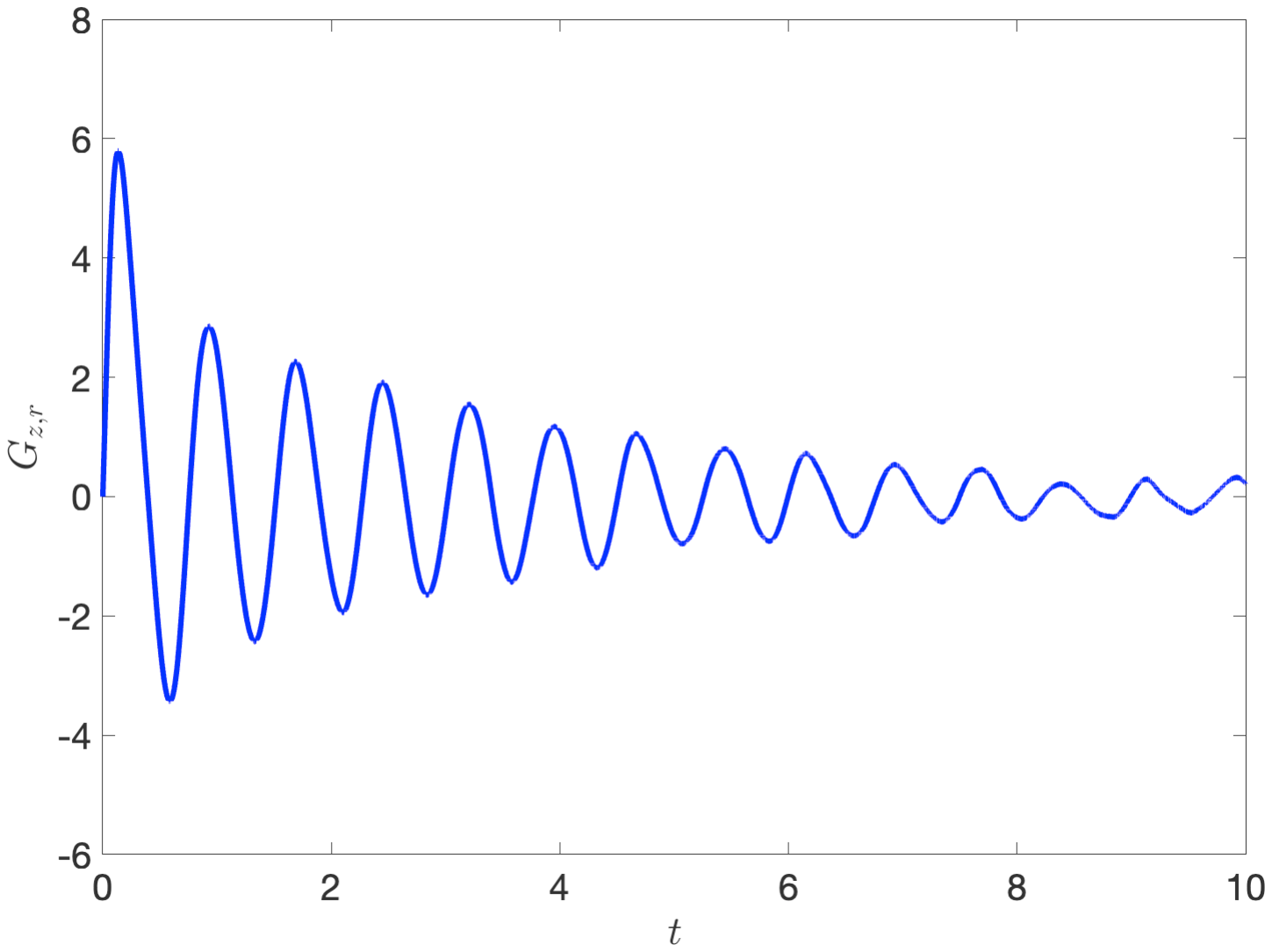}
        \label{fig:sub4}
    }
    
    \caption{Estimates of $G_{z,\rho}$ obtained using direct numerical simulation and different Koopman dictionaries.}
    \label{fig:Gz_estimates}
\end{figure*}

Figure \ref{fig:sub3} shows that using this methodology we need a lower number of dictionary functions to achieve an accurate representation of the long-time behaviour of the Green's functions, as the slowly decaying modes are also well-captured  when considering $P=14$ as highest polynomial order; see the rapid convergence of the eigenvalues with $P$ in Fig. \ref{fig:spectra_c}. Nonetheless, there is a price to be paid: as shown in the inset of Fig. \ref{fig:sub3}, the deep modes degrade the quality of the short-time behaviour of the Green's function (see also Eq. \ref{short}), as the gradient operation needed to construct $\Delta_i$ is not expected to be robust for poorly resolved modes.   Indeed, one can see in the inset that it is necessary to cut off the modes corresponding to $\mathcal{R}(\lambda_j)<-1.5$ in order to avoid excessive amplitude of oscillations for $t\leq1$. 


\section{Conclusions}\label{conclusions}

In this paper we have advanced the understanding of the link between forced and free fluctuations in nonequilibrium system by combining the formalism of response theory and Koopmanism. The key notion is that through Koopmanism it is possible to derive an interpretable form of the FDT whereby general response operators are expressed as a sum of terms, each associated with a specific mode of variability of the unperturbed system. It is possible to compute each of these terms once we choose a Koopman dictionary. Our efforts here bring together in a coherent framework earlier attempts in this direction \cite{Santos2022,Zagli2024,Lucarini2025}. 

We have tested the performance of different Koopman dictionaries of a prototypical nonequilibrium system, namely an SDE where the drift is given by the Lorenz '63 model with the classical value of the parameters and the noise is additive, isotropic, and non-degenerate.

A first set of dictionaries is  given by  characteristic functions of cells providing a  regular cubic tessellation of the 3-dimensional space with cubes of size $\mathrm{d}x$, with $\mathrm{d}x$= 1, 2, and 4.  Indeed, they lead to building Ulam partitions of the system of different level of granularity.  

A second option we follow amounts to a data-adaptive version of the previous construction, whereby the dictionary is given by characteristic functions of the Vorono\"i cells associated with a k-means clustering of the output of the model. We consider a cardinality of the clusters that approximately matches the number of cells visited by the system's trajectory in the regular gridding. 

Finally, we consider tensorized Chebyshev polynomials in the three variables $x$, $y$ ,$z$ up to a chosen order $P$. We consider $P$=14, 16, 18, and 20.  These dictionaries are fundamentally different from the previous ones because the considered functions are differentiable. 

While the methods employed are equivalent in analytical terms, the numerical and practical implementation is radically different. The first encouraging result is that in all considered cases we are able to obtain coherent estimates of the Koopman eigenvalues that show convergence as we consider a larger dictionary. Additionally, in all cases we are able to  accurately capture spectral features associated with key dynamical features of the deterministic component of  the system.

Such a convergence of the Koopman spectrum  with resolution is faster for dictionary 2. than for dictionary 1. Similarly, when considering comparable resolutions, the reconstruction of the response operator is better for dictionary 2 than for dictionary 1, which indicates the importance of using a data-adaptive method for selecting the Koopman dictionary. The use of characteristic functions leads us to represent the system as a finite-state Markov chain, which leads to extremely efficient and easily-implementable algebraic formulas for understanding virtually everything about the system at the chosen level of coarse-graining, see \cite{Lucarini2025}. The encouraging performance of the Markov chain representation derived from k-means clustering, which is relatively immune to the curse of dimensionality especially if one considers the bisecting k-means variant \cite{Souza2024}, is extremely promising in terms of supplementing Markov state modeling \cite{Laio2006,Pande2010,Bowman2014,Husic2018} with powerful tools for systematically studying the sensitivity, response, and criticality of complex systems. 

The role of fast decaying modes is clearer at short times, and an accurate evaluation of the Green’s function has proven a difficult task in this regime. This is especially true for the strategy used following dictionary 3., because of instabilities associated with the need to evaluate derivatives of rapidly varying dictionary functions. On the other hand, the use of tensorized Chebyshev polynomials proves extremely efficient in terms of capturing the long-term behaviour of the Green's function even when a limited dictionary is used. The numerical instability associated with short-time dynamics leads, however, to the need to preprocess the data using the SVD.  This angle could become of special relevance in a data-driven context if the observational data are first processed with SINDy \cite{Brunton2016Sindy} or its stochastic version \cite{Wanner2024} in order to derive an explicit dynamical model. 

Our exercise shows promising results, even if it is apparent that no easily recognisable silver bullet exists for deploying Koopmanism to predict the response of a complex system to perturbations. 

In the future, we will employ the methods developed here a) to derive reduced order response operators for high-dimensional systems; and b) to develop strategies for anticipating critical transitions by looking at the properties of the spectral gap in a reduced space. In the latter case, our goal is to make progress  beyond the standard early warning indicators based on critical slowing down and increased variance \cite{Lenton2012}, and to extend some recent results on the detection of critical behaviour \cite{Lohmann2025} in a functional analytical fashion. 

Additionally, following the promising results obtained using Koopman-based methods for studying climate variability \cite{Navarra2021,Navarra2024} and the theoretical framework proposed in \cite{LucariniChekroun2023,LucariniChekroun2024,Chekroun2024Kolmogorov}, we aim at deploying this technology to advance our understanding of the link between climate variability and climate change. 

Finally, following earlier promising results \cite{Lucarini2025}, we will pursue the explicit computation of nonlinear response operators in order to improve the quantitative prediction of a system's response to perturbations and to identify more complex interactions between internal feedbacks and acting forcings.

\subsection*{Acknowledgements}
V.L. acknowledges the partial support provided by the Horizon Europe Project ClimTIP (Grant No. 100018693) and by ARIA SCOP-PR01-P003 - Advancing Tipping Point Early Warning AdvanTip. V.L. and J.M. acknowledge the partial support provided by the CROPS RETF project funded by the University of Reading and by the EPSRC project LINK (Grant No. EP/Y026675/1). V.L. and M.S.G. acknowledge the partial support provided by the Horizon Europe project Past2Future (Grant No. 101184070). N.Z. has been supported by the Wallenberg Initiative on Networks and Quantum Information (WINQ).

\providecommand{\noopsort}[1]{}\providecommand{\singleletter}[1]{#1}%

\end{document}